\documentclass[
superscriptaddress,
preprintnumbers,
nofootinbib,
 amsmath,amssymb,
 aps,
]{revtex4-2}

\usepackage{amsfonts}
\usepackage{booktabs}
\usepackage{mathtools}
\usepackage{natbib}
\allowdisplaybreaks[4]
\usepackage{tensor}
\usepackage[dvipsnames]{xcolor}
\usepackage{IEEEtrantools}
\usepackage{upgreek}
\usepackage{graphicx}
\usepackage{caption}
\usepackage{float}
\usepackage{subcaption}
\usepackage[export]{adjustbox}
\usepackage{dcolumn}
\usepackage{bm}
\usepackage[dvipsnames]{xcolor}
\usepackage{hyperref}
\hypersetup{
	colorlinks =true,
	linkcolor =black,
	pdftitle={Renormalized holographic entanglement entropy for Quadratic Curvature Gravity},
	citecolor=MidnightBlue,
	filecolor=WildStrawberry,
	urlcolor=black,
}

\newcommand{\detG}{\sqrt{|\mathcal{G}|}}
\newcommand{\deth}{\sqrt{|h|}}
\newcommand{\detsigma}{\sqrt{|\sigma|}}
\newcommand{\detsigtil}{\sqrt{|\tilde{\sigma}|}}

\newcommand{\vol}{\text{Vol}}

\newcommand{\Leff}{L_{\text{eff}}}
\newcommand{\Lambdaeff}{\Lambda_\mathrm{eff}}

\newcommand{\IrenQG}{I^{\mathrm{ren}}_{\mathrm{QCG}}}

\newcommand{\SKt}{S_{\mathrm{KT}}}

\newcommand{\M}{{\mathcal{M}}}

\newcommand{\E}{\mathcal{E}}

\newcommand{\diff}{\mathrm{d}}
\newcommand{\diffdone}{{\diff^{d+1}}}
\newcommand{\diffmone}{{\diff^{d-1}}}

\newcommand{\QRicT}{R^\mu_\nu R^\nu_\mu}
\newcommand{\QRiem}{R^{\mu\nu}_{\kappa\lambda}R_{\mu\nu}^{\kappa\lambda}}

\DeclareMathOperator{\tr}{tr}

\begin{document}

\preprint{\href{https://arxiv.org/abs/2002.06111}{hep-th/2102.11242}}

\title{Renormalized holographic entanglement entropy for Quadratic Curvature Gravity}

\author{Giorgos Anastasiou}
\email{georgios.anastasiou@pucv.cl}
\affiliation{Instituto de F\'isica, Pontificia Universidad Cat\'olica de Valpara\'iso,\\ Casilla 4059, Valpara\'iso, Chile.}

\author{Ignacio J. Araya}
\email{araya.quezada.ignacio@gmail.com}
\affiliation{Instituto de Ciencias Exactas y Naturales, Facultad de Ciencias, Universidad Arturo Prat,\\ Avenida Arturo Prat Chac\'on 2120, 1110939, Iquique, Chile}

\author{Javier Moreno}
\email{francisco.moreno.g@mail.pucv.cl}
\affiliation{Instituto de F\'isica, Pontificia Universidad Cat\'olica de Valpara\'iso,\\ Casilla 4059, Valpara\'iso, Chile.}
\affiliation{Center for Quantum Mathematics and Physics (QMAP),\\
Department of Physics \& Astronomy, University of California, Davis, CA 95616 USA}

\author{Rodrigo Olea}
\email{rodrigo.olea@unab.cl}
\affiliation{Departamento de Ciencias F\'isicas, Universidad Andres Bello,\\
Sazi\'e 2212, Piso 7, Santiago, Chile}

\author{David Rivera-Betancour}
\email{david.rivera-betancour@zimbra.polytechnique.fr}
\affiliation{Centre de Physique Th\'eorique, CNRS, \'Ecole Polytechnique,\\*91128 Palaiseau Cedex, France.}

\begin{abstract}
\noindent We derive a covariant expression for the renormalized holographic entanglement entropy for Conformal Field Theories (CFTs) dual to Quadratic Curvature Gravity in arbitrary dimensions. This expression is written as the sum of the bare entanglement entropy functional obtained using standard conical defect techniques, and a counterterm defined at the boundary of the extremal surface of the functional. The latter corresponds to the cod-2 self-replicating part of the extrinsic counterterms when evaluated on the replica orbifold. This renormalization method isolates the universal terms of the holographic entanglement entropy functional. We use it to compute the standard $C$-function candidate for CFTs of arbitrary dimension, and the type-B anomaly coefficient $c$ for 4-dimensional CFTs.
\end{abstract}

\maketitle


\section{Introduction}

Higher-curvature gravity theories have attracted considerable attention in the literature for various reasons. This is mainly due to their better  ultraviolet (UV) completion properties \cite{Stelle:1976gc,Capper:1975ig,tHooft:1974toh,Deser:1974cz,Deser:1974xq}, what provides a promising scenario for deciphering quantum gravity properties. Indeed, the low energy effective action of String Theory is characterized by the presence of a series of higher-derivative terms added on top of the Einstein-Hilbert action.

Quadratic Curvature Gravity (QCG) is one of the simplest examples of a higher-curvature gravitational theory. Its action is given by 
\begin{equation}\label{QG2}
I_{\mathrm{QCG}}=\frac{1}{16 \pi G_{N}}\int\limits_{\M}\diffdone x\detG\left(R-2\Lambda_0+\alpha R^2+\beta R^{\mu}_{\nu}R^{\nu}_{\mu}+\gamma GB\right) \,,
\end{equation}
where the \emph{bare} cosmological constant is $\Lambda_0=-\frac{d(d-1)}{2L_0^{2}}$, in terms of the radius of AdS space $L_0$. Here, the scalar density $GB$ stands for the Gauss-Bonnet term, which reads $GB=R^2-4R_{\mu}^{\nu}R_{\nu}^{\mu}+R^{\mu \nu}_{\kappa \lambda} R^{\kappa \lambda}_{\mu \nu}$. The equation of motion (EOM) of the theory \eqref{QG2} is fourth-order in differential terms of the metric $\mathcal{G}_{\mu\nu}$, when arbitrary couplings $(\alpha,\beta,\gamma)$ are considered. However, if $\alpha=\beta=0$, i.e., the Gauss-Bonnet term is the only modification considered, the EOM becomes second order \cite{Lanczos:1938sf}, as the theory goes back to the Lovelock class \cite{lovelock1970divergence,lovelock1971einstein}.

A larger family is that of Generalized Quasi-topological Gravities (GQTGs) that have sparked a lot of interest in recent years \cite{Myers:2010ru,Oliva:2010eb,Bueno:2016xff,Hennigar:2017ego,Bueno:2019ltp,Bueno:2019ycr}. Its main feature is the fact that the EOM is reduced to second order when evaluated in a static and spherically-symmetric ansatz.

In $d+1=4$ dimensions, of particular importance is the point of the parametric space of QCG where $\alpha=-1/(2\Lambda_0)$ and $\beta=3/(2\Lambda_0)$. In this dimension, $\gamma$ plays no role as the Gauss-Bonnet term is not dynamical. It corresponds to Critical Gravity, introduced in ref.\cite{Lu:2011zk}, a theory which becomes trivial when evaluated on Einstein spacetimes \cite{Anastasiou:2017rjf}. Interestingly enough, for relaxed AdS asymptotic conditions, this theory constitutes the gravitational dual of three-dimensional logarithmic CFT.

The Holographic Principle has a concrete realization in the form of the anti-de Sitter/Conformal Field Theory (AdS/CFT) correspondence introduced in refs.\cite{Maldacena:1997re,Gubser:1998bc,Witten:1998qj}. In this context, higher-curvature gravity theories have been the subject of extensive study as toy models due to the appealing properties of their dual field theories. This class of theories probes a broader range of CFTs than standard Einstein gravity. Indeed, the type-A and type-B anomalies of their four-dimensional holographic counterparts do not coincide, unlike in the Einstein case \cite{Nojiri:1999mh,Ghodsi:2019xrx}. Furthermore, the couplings of higher curvature theories modify the ratio of shear viscosity $\eta$ to entropy density $s$ of their hydrodynamic duals, with respect to the value provided by Einstein gravity of $\eta / s = 1/(4\pi)$ \cite{Policastro:2001yc,Buchel:2003tz,Kovtun:2004de}, as can be seen in five-dimensional Einstein-Gauss-Bonnet theory \cite{Brigante:2007nu}. This indicates that the Kovtun-Son-Starinets (KSS) bound is not universal in the sense that its value has explicit dependence on the higher curvature couplings of the gravity theory. However, as causality considerations further constrain said couplings \cite{Camanho:2009vw}, it is possible to obtain modified KSS bounds applicable to classes of higher curvature theories \cite{Brigante:2008gz,deBoer:2009pn,Camanho:2010ru}. This is another example of how the presence of higher curvature terms can break degeneracies in holographic quantities, helping to pinpoint the holographic dictionary more precisely.
    
A similar non-trivial behavior is present in the universal terms of the Entanglement Entropy (EE) for CFTs dual to higher-curvature gravity theories \cite{Myers:2010tj,Myers:2010xs,Mezei:2014zla,Bueno:2018yzo,Bueno:2020odt,Hung:2011xb}. EE appears as a useful tool to explore aspects of quantum entanglement in strongly coupled field theories. In a quantum model described by the density matrix $\varrho$, the EE between a subsystem $A$ and its complement $A^c$ is given by the von Neumann entropy of the reduced matrix $\varrho_A = \tr_{A^c} \varrho$, 

\begin{equation} \label{EEdef}
S_A = - \tr_A \left(\varrho_A \text{log} \varrho_A\right) \,.
\end{equation}
As many observables in field theory, EE is a UV divergent measure. Its main characteristic is the fact that the leading divergence always scales as the cod-2 area of the entangling surface $\partial A$ that separates the subsystem $A$ with its complement, independently of its shape\footnote{This assumes local field theories and continuity of the fields and their derivatives across the entangling surface.}. This is due to the UV degrees of freedom which are localized at the vicinity of the entangling surface \cite{Grover:2011fa,Myers:2012ed,Liu:2012eea}.

Of great interest are the universal features associated to the Renormalization Group (RG) flow of the Quantum Field Theory (QFT), in the expansion of the EE near $\partial A$. They correspond to scheme-independent terms that probe the number of effective degrees of freedom of the respective theory. In odd dimensions, it is the finite part of the EE that can be related to the $F$-function that is monotonic along the RG flow \cite{Jafferis:2011zi,Klebanov:2011gs,Casini:2011kv}. In even-dimensions, one can identify the type-A and type-B anomaly coefficients of the corresponding field theory, from the logarithmic term of the EE expansion. As the type-A anomaly monotonically decreases when going from ultraviolet to infrared (IR), it is a good $C$-function candidate \cite{Zamolodchikov:1986gt,Cardy:1988cwa,Komargodski:2011vj}.

In principle, the calculation of the EE in QFT is rather involved. Using the real-time formalism or  the replica-trick in the path-integral quantization demands analytic techniques that can be applied only for particular shapes of $A$, in low enough dimensions \cite{Calabrese:2004eu}.

Major progress was made in this direction by considering the AdS/CFT duality. Ryu and Takayanagi (RT) conjectured that EE is given by a generalization of the Bekenstein-Hawking formula for cod-2 surfaces which are not the fixed point of a continuous isometry \cite{Ryu:2006bv,Ryu:2006ef}. More specifically, the EE of a spatial subregion $A$ is given by the area of a minimal homologous cod-2 surface $\Sigma$, i.e.,
\begin{equation} \label{RT}
S_A = \frac{\text{Area} \left(\Sigma\right)}{4G_N} \,.
\end{equation}
The surface is embedded in the bulk solution of the dual gravity theory and anchored at $\partial A$.
 
The proof behind the conjectured RT formula \eqref{RT}, came a few years later by Lewkowycz and Maldacena \cite{Lewkowycz:2013nqa}. The central idea is the relation between holographic entanglement entropy (HEE) and Euclidean gravity action, in the saddle-point approximation of AdS/CFT. This is achieved by constructing bulk solutions from a set of boundary conditions that are not $U(1)$ invariant. As a consequence, one may extend the replica symmetry, needed for the calculation of the EE, to the bulk. In particular, the HEE is given by the limit
\begin{equation}\label{eq:SLM}
S_{\text{EE}}=-\lim\limits_{\vartheta\rightarrow1}\partial_{\vartheta}I_{E}\left[
\mathcal{M}^{(\vartheta)}\right] \,,
\end{equation}
where $I_{E}\left[
\mathcal{M}^{(\vartheta)}\right]$ denotes the Euclidean gravity action evaluated on a conically-singular manifold. The orbifold $\mathcal{M}^{(\vartheta)}$  is constructed as the bulk gravity dual of the replica CFT obtained through the standard replica trick in \cite{Lewkowycz:2013nqa,Nishioka:2018khk,Fursaev:2013fta}. As the replica symmetry is discrete, $\mathcal{M}^{(\vartheta)}$ is a squashed-cone (having no $U(1)$ isometry), and it has an angular deficit of $2\pi \left(1- \vartheta\right)$, such that $\vartheta=\frac{1}{m}$ is related to the replica index $m$. 

As pointed out in ref.\cite{Dong:2016fnf}, in the case of HEE in Einstein-AdS gravity, $\frac{\left(1-\vartheta\right)}{4G_{N}}$ can be identified with the tension of a cosmic brane coupled to the ambient geometry through the Nambu-Goto action. Thus, in the tensionless limit ($\vartheta \rightarrow 1$), the area of the cosmic brane anchored at  the boundary gives the HEE, trivially recovering the RT prescription \cite{Ryu:2006bv}. However, eq.\eqref{eq:SLM} also holds for generic gravity theories, such that upon evaluating the action on the $\mathcal{M}^{(\vartheta)}$ orbifold, the correct entropy functional emerges naturally. The most notable contributions in this direction have been given by Dong \cite{Dong:2013qoa}, Camps \cite{Camps:2013zua} and Miao \cite{Miao:2014nxa}. In the case of QCG, the evaluation of the arbitrary quadratic curvature invariants on squashed-cones was firstly done by Fursaev, Patrushev and Solodukhin (FPS) in ref.\cite{Fursaev:2013fta} using distributional geometry techniques. In the present work we exploit the fact that, as the HEE functional depends only on the gravity action under consideration, its renormalization is inherited from the renormalization of the bulk gravity action \cite{Taylor:2016aoi}.

On previous papers, based on a renormalization scheme defined by the addition of extrinsic counterterms,  we showed that for odd-dimensional CFTs dual to Einstein-AdS gravity, the finite part of the HEE is the sum of the Euler characteristic of the RT surface, and a curvature term on that surface \cite{Anastasiou:2018rla}\footnote{In ref.\cite{Taylor:2020uwf}, the authors arrive at the same conclusion using standard Holographic Renormalization techniques.}. The coefficient of the Euler characteristic is matched to the $F$-term, a quantity that has been shown to be monotonic along RG flows \cite{Myers:2010tj,Nishioka:2018khk,Klebanov:2011gs,Jafferis:2011zi,Myers:2010xs,Casini:2012ei}. In particular, in 3-dimensional CFTs,  the $F$-quantity is robust under continuous deformations of the entangling region \cite{Anastasiou:2020smm}. In the case of even-dimensional CFTs, the $C$-function candidate is identified as the coefficient of the log part of the EE. As shown in ref.\cite{Anastasiou:2019ldc}, this universal contribution can be isolated when the HEE is written in terms of a covariant functional which is the usual area term plus a cod-3 extrinsic boundary counterterm. The latter term cancels power-law divergences in the HEE formula. 

In this paper, we turn our interest to the case of QCG. There, we will explicitly show that the same form of  extrinsic counterterms (only with a different coupling constant) is able to isolate the universal part of the HEE for CFTs dual to this type of higher curvature gravity theories. This can be achieved by considering particular shapes of entangling regions (i.e., spheres and cylinders), where the $C$-function candidates and other holographic quantities like the type-B anomaly coefficient $c$ can be directly obtained \cite{Henningson:1998gx,Nojiri:1999mh,Ghodsi:2019xrx}.

\section{Counterterms of a different sort: Kounterterms.}

In General Relativity, it is straightforward to prove that second derivatives in the normal coordinate appear in the Lagrangian. One may think of an analogous
system in Classical Mechanics: a Lagrangian with a linear dependence on the acceleration $\ddot{q}$, of the type $L(q,\dot{q},\ddot{q})=\ddot{q}h(q)+\ell(q,\dot{q})$, where $q$ is the generalized coordinate. An arbitrary variation of this action gives rise to a field equation which is still of second order in the time derivative, while producing a term that contains both $\delta q$ and $\delta \dot{q}$ at the boundary. The addition of a total derivative to $L$ leads to a surface term where only  $\delta q$ appears, what is a feature
of first-order Lagrangians.
The above reasoning can be mimicked in gravity, what implies that the bulk action
has to be augmented by the Gibbons-Hawking term at the boundary, in order to ensure a well-defined action principle
for a Dirichlet boundary condition on the boundary metric $h_{ab}$ \cite{Gibbons:1976ue}. The introduction of a cosmological constant does not
modify the surface terms of the theory, but it does change the asymptotic behavior of the metric, such that the canonical 
momentum is no longer finite at the boundary.
In the early days of the AdS/CFT correspondence, the renormalization  of AdS gravity was achieved by the addition
of intrinsic counterterms, such that the Dirichlet boundary condition was not spoilt \cite{Emparan:1999pm,Balasubramanian:1999re,Kraus:1999di}.
It is then that, by a proper rescaling of the metric $h_{ab}$, holographic correlation functions are obtained as variations with respect to
the source at the conformal boundary $g_{(0)ab}$.

However, it was later pointed out  by Papadimitriou and Skenderis \cite{Papadimitriou:2005ii}
that a Dirichlet condition on the boundary metric $h_{ab}$ does not make sense in asymptotically AdS (AAdS) spaces. As a matter of fact, the asymptotic expansion of this field $h_{ab}=\frac{g_{(0)ab}}{z^2}+...$ reveals an infinite conformal factor near the boundary. As a consequence, the only way to have a well-posed variational principle in AdS gravity is fixing --instead-- the metric
$g_{(0)ab}$ at the conformal boundary.
This argument implies that the addition of counterterms is required not only for cancelling divergent terms in the variation of the action, but also for the consistency of the variational problem on $g_{(0)ab}$.

Along the same line, we stress the fact that there is a blissful accident in AdS gravity: the leading order in the asymptotic expansion of
the extrinsic curvature is (up to a numerical factor) the same as the one in $h_{ab}$. Indeed, in the Fefferman-Graham frame, $K_{ab}=\frac{1}{\ell}\frac{g_{(0)ab}}{z^2}+...$, what has been recently emphasized in ref.\cite{Witten:2018lgb}. This simple observation means that one can express
also variations of $K_{ab}$ in terms of variations of $g_{(0)ab}$. This also implies that one may consider surface terms which depend on the
extrinsic curvature and act as counterterms, in the sense that they cancel divergent contributions in the AdS gravity action.
So, even though they are plain incompatible with a Dirichlet condition for the full boundary metric $h_{ab}$, they can still
reproduce the correct holographic stress tensor varying with respect to $g_{(0)ab}$.

The above reasoning, which opens the possibility to look for an alternative sort of counterterms, is
justified by the lack of a closed expression  for the series in arbitrary dimensions.
More than twenty years ago, evidence was provided on the fact that topological terms were able to regulate the
variation of the AdS gravity action in even dimensions \cite{Aros:1999id,Aros:1999kt}, though based on the study
of particular solutions.
As for the Euclidean action, the addition of the Euler term at the boundary of $d=2n$ dimensions renders it finite in AAdS solutions
if the coupling is adequately chosen \cite{Olea:2005gb}. However, it was not clear what this prescription to
renormalize AdS gravity had to do with Holographic Renormalization and the addition of standard counterterms.

A first step towards the understanding of this issue was given in ref.\cite{Olea:2005gb}, where topological terms in the bulk are equivalently written as the corresponding Chern form $B_{2n-1}$ at the boundary,
\begin{equation} \label{Bdodd}
B_{2n-1}=-2n\int\limits_0^1\diff s\deth\updelta^{a_1\cdots a_{2n-1}}_{b_1\cdots b_{2n-1}}K^{b_1}_{a_1} \left(\frac{1}{2}\mathcal{R}^{b_2b_3}_{a_2a_3}-s^2K^{b_2}_{a_2}K^{b_3}_{a_3}\right)\times\ldots
\times\left(\frac{1}{2}\mathcal{R}^{b_{2n-2}b_{2n-1}}_{a_{2n-2}a_{2n-1}}-s^2K^{b_{2n-2}}_{a_{{2n-2}}}K^{b_{2n-1}}_{a_{2n-1}}\right) \,.
\end{equation}
For the first time, counterterms which depend on $K_{ab}$ were proposed to deal with the renormalization of AdS gravity. Here, $h_{ab}=g_{ab} \left(z,x\right)/z^2$ is the induced metric at constant Poincar\'e coordinate $z$, $\mathcal{R}_{cd}^{ab}$ is the intrinsic Riemann curvature tensor, $K_{ab}$ is the extrinsic curvature and $\updelta^{a_1\cdots a_{2n-1}}_{b_1\cdots b_{2n-1}}$ is the generalized Kronecker delta. In this notation, the indices of the generalized Kronecker delta are contracted with those of the rest of the tensors in the integrand.

A similar structure at the boundary of $d+1=2n+1$ dimensions was far more difficult to obtain. In particular,
due to the fact that there is no equivalent form in the bulk for such boundary term. The extensive use
of field-theory tools in the context of anomalies (Chern-Simons and transgression forms, homotopy operator, etc.)
allows to make a concrete proposal for that case. The resulting term, in essence, shares common
properties with the Chern form, as it is a given polynomial of the extrinsic and intrinsic curvatures \cite{Olea:2006vd},
but  it did not exist in the mathematical literature before. Its specific form is given by the following expression

\begin{align}\label{eq:Bdeven}
B_{2n}=-2n \int \limits_0^1 \diff s\int\limits_0^s\diff t\deth\,\updelta_{b_1\cdots b_{2n-1}}^{a_1\cdots a_{2n-1}} K^{b_1}_{a_1} \left(\frac{1}{2} \mathcal{R}^{b_2 b_3}_{a_2 a_3}-s^2 K^{b_2}_{a_2} K^{b_3}_{a_3} + \frac{t^2}{\Leff^2} \updelta^{b_2}_{a_2} \updelta^{b_3}_{a_3} \right)\times\cdots \nonumber \\
\cdots\times\left(\frac{1}{2}\mathcal{R}^{b_{2n-2}b_{2n-1}}_{a_{2n-2}a_{2n-1}}-s^2K^{b_{2n-2}}_{a_{2n-2}}K^{b_{2n-1}}_{a_{2n-1}}+\frac{t^2}{\Leff^2}\updelta^{b_{2n-2}}_{a_{2n-2}}\updelta^{b_{2n-1}}_{a_{2n-1}}\right) \,,
\end{align}
where $\Leff$ is the effective AdS radius of the theory. 

One may think that the renormalization procedure described above, dubbed \emph{Kounterterms}, may lead to a variational principle which is at odds with the holographic description of AdS gravity in terms of the boundary source $g_{(0)ab}$, as it seems to require a different boundary condition on the extrinsic curvature. But the analysis portrayed above gives a firmer ground
to the addition of Kounterterms to the gravitational action: the total action is consistent with
a holographic description, as its variation is both finite and given in terms of $\delta g_{(0)ab}$.

This simple reasoning suggests the resummation of the counterterm series as an expression in terms of
the extrinsic curvature. As a matter of fact, an asymptotic expansion of the term $B_{d}$ reproduces
the counterterms, once the Gibbons-Hawking is correctly isolated. This was first sketched in ref.\cite{Miskovic:2009bm}
and analyzed in greater detail in a recent paper \cite{Anastasiou:2020zwc}\footnote{In ref.\cite{Anastasiou:2020zwc}, it  was shown that, in Einstein-AdS gravity, the Kounterterms are the resummation of the counterterms for asymptotically conformally flat (ACF) manifolds in arbitrary dimensions. 
For a generic AAdS space, there is a mismatch between counterterms and Kounterterms, consisting on terms which are the dimensional continuation of conformal invariants at the boundary. At the lowest order, this difference is a Weyl-squared term, which is identically vanishing for a conformally flat boundary. However, it may be the case that this condition is relaxed by taking Weyl$^{2}=0$ instead, what would be the analogue of demanding local flatness ($Rie=0)$ \emph{vs} a vanishing Kretschmann scalar for a given spacetime.
}. In addition, earlier works in the mathematical literature \cite{2000math.....11051A,2005math......4161A}, indicate that the Chern form is fundamental in defining the renormalized volume of an Einstein space.

The Kounterterm method  has been used to deal with the construction of conserved quantities and the thermodynamic description
of black holes in Einstein-Gauss-Bonnet AdS and, in general, Lovelock AdS gravity. Furthermore, it has also linked
the concept of Conformal Mass to the addition of boundary terms in Einstein-Hilbert \cite{Jatkar:2014npa} and higher-curvature gravity \cite{Arenas-Henriquez:2017xnr,Arenas-Henriquez:2019rph}. 
Evidence has been given that Kounterterms can provide finite conserved charges in QCG, as well \cite{Giribet:2018hck,Giribet:2020aks,GMOPodd}. 

In the present paper, we extensively use its properties to deal with the problem
of renormalization of HEE in QCG.

\section{Renormalized EE from the replica orbifold}

In this section, we compute the renormalized HEE in CFTs dual to QCG. In order to determine the EE functional from the bulk gravity Lagrangian we use the results by Fursaev, Patrushev and Solodukhin for the evaluation of quadratic curvature invariants in conically singular manifolds \cite{Fursaev:2013fta}. The renormalization of HEE is then inherited from the renormalization of the bulk action by the addition of cod-1 Kounterterms and derived from a set of cod-2 relations with respect to the bulk and boundary dimensions. In particular, it is a remarkable property of the Kounterterm $B_{d}$ --when evaluated on orbifolds-- that the singular part gives rise to the same structure in two dimensions lower $\left(B_{d-2}\right)$. This self-replicating feature of $B_{d}$ induces cod-3 Kounterterms on $\partial \Sigma$ which renormalize the cod-2 FPS functional. Based on this construction, we determine the universal terms of the HEE for this theory.

\subsection{Curvature invariants on the cone}

Now, we consider the expressions for the quadratic curvature invariants
evaluated on squashed cones (without U(1) isometry), as given in refs.\cite{Fursaev:2013fta,Anastasiou:2017xjr}. Following this procedure, we have
\begin{IEEEeqnarray}{cl} 
\int\limits_{\mathcal{M}^{(\vartheta)}}&\diffdone x\detG\ R^2=\int\limits_{\mathcal{M}^{(\vartheta)}\setminus\Sigma}\diffdone x\detG\ R^2  \label{RicciScal2Cone} \\
&+8\pi(1-\vartheta)\int\limits_\Sigma \diffmone y\detsigma\left(\tilde{R}
+2R^{(i)(i)}-R^{(i)(j)(i)(j)}-2{\mathcal{K}^{(i)}}^{\bar{\mu}}_{[\bar{\mu}}{\mathcal{K}^{(i)}}^{\bar{\nu}}_{\bar{\nu}]}\right)+\ldots\nonumber \,,\\
\int\limits_{\mathcal{M}^{(\vartheta)}}\diffdone x&\detG\ \QRicT=\int\limits_{\mathcal{M}^{(\vartheta)}\setminus\Sigma}\diffdone x\detG\ \QRicT \label{Ricci2Cone}\\
&+4\pi(1-\vartheta)\int\limits_\Sigma \diffmone y\detsigma\left(R^{(i)(i)}-\frac{1}{2}{\mathcal{K}^{(i)}}^{\bar{\mu}}_{\bar{\mu}}{\mathcal{K}^{(i)}}^{\bar{\nu}}_{\bar{\nu}}\right)+\ldots\nonumber\,,\\
\int\limits_{\mathcal{M}^{(\vartheta)}}\diffdone x&\detG\ \QRiem=\int\limits_{\mathcal{M}^{(\vartheta)}\setminus\Sigma}\diffdone x\detG\ \QRiem \label{Riem2Cone}\\
&+8\pi(1-\vartheta)\int\limits_\Sigma \diffmone y\detsigma\left(R^{(i)(j)(i)(j)}-{\mathcal{K}^{(i)}}^{\bar{\nu}}_{\bar{\mu}}{\mathcal{K}^{(i)}}^{\bar{\mu}}_{\bar{\nu}}\right)+\ldots\nonumber \,,
\end{IEEEeqnarray}
where the ellipsis denotes terms of higher order in $\left(  1-\vartheta\right)  $, $\Sigma$ is the fixed-point
set of the replica symmetry (i.e., the cod-2 surface at the apex of the cone),
$i,j$ are indices normal to $\Sigma$, $\bar{\mu},\bar{\nu}$ are indices in the worldvolume of
$\Sigma$, ${K^{(i)}}_{\bar{\mu}\bar{\nu}}$ is the extrinsic curvature of the
cod-2 foliation along the $i$ direction, and summation is implied on repeated
$i,j$ indices.

We also consider the evaluation of  the Ricci scalar
and the GB term on the squashed cone, following the results of ref.\cite{Fursaev:2013fta}. We therefore have
\begin{IEEEeqnarray}{cl}
\int\limits_{\mathcal{M}^{(\vartheta)}}\diffdone x\detG\ R&=\int\limits_{\mathcal{M}^{(\vartheta)}\setminus\Sigma}\diffdone x\detG\ R+4\pi(1-\vartheta)\int\limits_\Sigma\diffmone y\detsigma \,,\\
\int\limits_{\mathcal{M}^{(\vartheta)}}\diffdone x\detG\ \text{GB}&=\int\limits_{\mathcal{M}^{(\vartheta)}\setminus\Sigma}\diffdone x\detG\ \text{GB}+8\pi(1-\vartheta)\int\limits_\Sigma\diffmone y\detsigma\ \tilde{R} \,.
\end{IEEEeqnarray}

\subsection{Extrinsic counterterms on the cone}


We begin our analysis of Kounterterms evaluated on squashed cones by considering the topological origin of the Chern form. When $\mathcal{M}$ is even dimensional, the existence of extrinsic counterterms $B_{2n-1}$ is guaranteed by the Euler theorem 
\begin{equation} \label{Euler}
\int \limits_{\mathcal{M}}
\diff^{2n}x \,\mathcal{E}_{2n}
= (4\pi)^{n} n! \chi(\mathcal{M}) +\int\limits_{\partial\mathcal{M}}
\diff^{2n-1}X\, B_{2n-1}\,,
\end{equation}
which unveils the dynamic equivalence to the topological term in the bulk 
\begin{equation}
\E_{2n}=\frac{\detG}{2^n}\updelta_{\mu_1\ldots\mu_{2n}}^{\nu_1\ldots\nu_{2n}}R^{\mu_1\mu_2}_{\nu_1\nu_2}\cdots R^{\mu_{2n-1}\mu_{2n}}_{\nu_{2n-1}\nu_{2n}}  \,.
\end{equation}
The above relation also identifies $B_{2n-1}$ with the corresponding Chern form, which appears as the correction to the Euler characteristic in a manifold with boundary.

For the purposes of the current study, we highlight the fact that the equivalence between bulk and boundary terms can be generalized to manifolds with conical deficits. 
Indeed, any Euler term $ \E_{2n}$ self-replicates in cod-2, that is,
\begin{equation} \label{Lovelockorbifold}
\int\limits_{\mathcal{M}^{(\vartheta)}}\diff^{2n}x \, \E_{2n}=\int\limits_{\mathcal{M}^{(\vartheta)}\setminus\Sigma}\diff^{2n}x  \, \E_{2n}+4 \pi n (1-\vartheta)\int\limits_\Sigma \diff^{2n-2}y \, \E_{2n-2} \,,
\end{equation}
when constructed on an orbifold $M^{(\vartheta)}$ \cite{Kastikainen:2020auf,Anastasiou:2019ldc}. This fact implies that the self-replication property extends to the Chern form, as well \cite{Anastasiou:2018mfk,Anastasiou:2018rla}

\begin{equation} \label{Kountereroddorbifold}
\int\limits_{\partial\mathcal{M}^{(\vartheta)}} \diff^{2n-1}X B_{2n-1}=\int\limits_{\partial\mathcal{M}^{(\vartheta)}\setminus\partial\Sigma} \diff^{2n-1}X B_{2n-1}+4\pi n (1-\vartheta)\int\limits_{\partial\Sigma} \diff^{2n-3}Y \, B_{2n-3} \,.
\end{equation}
Unlike the Chern form, Kounterterms $B_{2n}$ for odd bulk dimensions are not associated to the Euler density. Therefore, the previous analysis for orbifolds cannot be repeated verbatim. However, even though $B_{2n-1}$ and $B_{2n}$ are different geometrical objects, it can be shown that, in presence of squashed conical singularities,
\begin{equation} \label{Kounterevenorbifold}
\int\limits_{\partial\mathcal{M}^{(\vartheta)}} \diff^{d}X B_{d}=\int\limits_{\partial\mathcal{M}^{(\vartheta)}\setminus\partial\Sigma} \diff^{d}X B_{d}+4\pi  \left\lfloor\frac{d+1}{2}\right\rfloor (1-\vartheta)\int\limits_{\partial\Sigma} \diff^{d-2}Y B_{d-2} \,,
\end{equation}
for either odd or even boundary dimension $d$  \cite{Anastasiou:2019ldc}. 
 
The extrinsic counterterms $B_{d-2}$ (living in cod-3 respect to the bulk space) cancel the divergences coming from the set of points anchoring the surface $\Sigma$ to the conformal boundary. It has been shown that they correctly isolate the universal terms of the entanglement and R\'enyi entropies for CFTs which are dual to Einstein gravity \cite{Anastasiou:2017xjr,Anastasiou:2018mfk,Anastasiou:2018rla,Anastasiou:2019ldc}. In that situation, the relation between EE and the notion of renormalized area was made manifest. 

In what follows, we seek to extend the renormalization scheme for cod-2 surfaces to one of the simplest examples of a higher-derivative gravity, such as QCG.

\subsection{The Kounterterm-renormalized QCG action}

We start by considering the generic renormalized QCG action \cite{Giribet:2020aks}
\begin{equation} \label{renQCG}
\IrenQG=I_{\mathrm{QCG}}+\frac{c_d}{16 \pi G_{N}}\int\limits_{\partial \M}B_{d} \,,
\end{equation}
where $B_{d}$ is the boundary Kounterterm defined in eq.\eqref{Kounterevenorbifold}. 

By solving the usual characteristic equation to find the vacuum of the theory, we obtain
the following relation for the effective cosmological constant and its
corresponding effective AdS radius
\begin{equation}
-\frac{1}{2\Lambdaeff}+\frac{\Lambda_{0}}{2
\Lambdaeff^{2}}   =\frac{\left(  d-3\right)  }{\left(d-1\right)
^{2}}\left[  \left(  d+1\right)  \alpha+\beta\right]  +\frac{\left(
d-3\right)  \left(  d-2\right)  }{d\left(  d-1\right)  }\gamma \,,
\end{equation}
where  $\Lambdaeff=-\frac{d\left(  d-1\right)  }{2\Leff^2}$. Also, the coupling $c_{d}$ of the boundary Kounterterm $B_{d}$ is fixed by
requiring the action of the pure AdS solution (vacuum) to be finite, thus
finding
\begin{equation}\label{eq:cd}
c_{d}=\begin{dcases*}a_d\frac{\left(  -1\right)  ^{\frac{d+1}{2}}\Leff^{d-1}}
{\left(  \frac{d+1}{2}\right)  \left(  d-1\right)  !}& \text{if $d$ odd \,,} \\
a_d\frac{\left(  -1\right)  ^{\frac{d}{2}}\Leff^{d-2}}
{2^{d-3}d\left(\left(\frac{d}{2}-1\right)  !\right)^2}&  \text{if $d$ even  \,,}\end{dcases*} 
\end{equation}
where the auxiliary function $a_d$ reads
\begin{equation}\label{eq:ad} a_d=1-\frac
{2d}{\Leff^2}\left[  \left(  d+1\right)  \alpha
+\beta+\frac{\left(d-2\right)\left(d-1\right)}{d}\gamma\right] \,.
\end{equation}
Based on the preceding relations, we evaluate the QCG action augmented with the boundary term \eqref{renQCG} on the squashed cone, obtaining%
\begin{equation}
\IrenQG\left[  M^{(\vartheta)}\right]  =\IrenQG\left[
M^{(\vartheta)}\backslash\partial\Sigma\right]  +\left(  1-\vartheta\right)
S^{\text{Univ}}_{\text{QCG}}+\ldots \,.
\end{equation}
When considering the Lewkowycz-Maldacena prescription \eqref{eq:SLM}, applied to the renormalized action \eqref{renQCG}, $S^{\text{Univ}}_{\text{QCG}}$ corresponds to the universal part of the HEE for CFTs dual to QCG, which is finite for odd-$d$ and log-divergent for even-$d$. Explicitly, the
$S^{\text{Univ}}_{\text{QCG}}$ functional is given by
\begin{IEEEeqnarray}{cl}\label{SrenQCG}
S^{\text{Univ}}_{\text{QCG}}=S_{\text{QCG}} + \SKt \,,
\end{IEEEeqnarray}
where
\begin{IEEEeqnarray}{cl}
S_{\text{QCG}}=&\frac{1}{4G_{N}}\Bigg\{\mathrm{Area}\left[\Sigma\right]+\int\limits_\Sigma\diffmone y\detsigma\bigg[2\alpha\left(\tilde{R}
+2R^{(i)(i)}-R^{(i)(j)(i)(j)}-2{\mathcal{K}^{(i)}}^{\bar{\mu}}_{[\bar{\mu}}{\mathcal{K}^{(i)}}^{\bar{\nu}}_{\bar{\nu}]}\right)\nonumber\\
&+\beta\left(R^{(i)(i)}-\frac{1}{2}{\mathcal{K}^{(i)}}^{\bar{\mu}}_{\bar{\mu}}{\mathcal{K}^{(i)}}^{\bar{\nu}}_{\bar{\nu}}\right)+2\gamma \tilde{R}\bigg]\Bigg\}\label{eq:SQCGbare} \,,
\end{IEEEeqnarray}
and the corresponding curvature terms are defined after eqs.(\ref{RicciScal2Cone},\ref{Ricci2Cone},\ref{Riem2Cone}). Alternatively, $S_{\text{QCG}}$ can be derived either following the prescription provided by Dong in ref.\cite{Dong:2013qoa} or the one by Camps in ref.\cite{Camps:2013zua}\footnote{Furthermore, in ref.\cite{Mozaffar:2016hmg}, $S_{\text{QCG}}$ was computed using field redefinition in the gravity theory.}. For gravity theories with curvature terms of cubic order or higher, these procedures give different HEE functionals, which is referred to as the \emph{splitting problem} \cite{Miao:2014nxa,Camps:2014voa,Miao:2015iba,Camps:2016gfs}. 

On the other hand, the Kounterterm obtains the form
\begin{equation}\label{eq:SKt}
\SKt=\frac{c_d}{4G_{N}}\left \lfloor{\frac{(d+1)}{2}}\right \rfloor\int\limits_{\partial\Sigma}B_{d-2} \,.
\end{equation}
For this expression, the cod-2 boundary term reads
\begin{equation}\label{eq:Bdminus2}
    B_{d-2}=
        \begin{cases}
        \begin{split}
   -(d-1)\int\limits_0^1\diff s\detsigtil&\updelta^{\bar{a}_1\cdots \bar{a}_{d-2}}_{\bar{b}_1\cdots \bar{b}_{d-2}}k^{\bar{b}_1}_{\bar{a}_1}\left(\frac{1}{2}\tilde{\mathcal{R}}^{\bar{b}_2\bar{b}_3}_{\bar{a}_2\bar{a}_3}-s^2k^{\bar{b}_2}_{\bar{a}_2}k^{\bar{b}_3}_{\bar{a}_3}\right)\times\cdots\\
    & \cdots\times\left(\frac{1}{2}\tilde{\mathcal{R}}^{\bar{b}_{d-3}\bar{b}_{d-2}}_{\bar{a}_{d-3}\bar{a}_{d-2}}-s^2k^{\bar{b}_{d-3}}_{\bar{a}_{d-3}}k^{\bar{b}_{d-2}}_{\bar{a}_{d-2}}\right) \,,
\end{split}& \text{odd }d \,,\\
\begin{split}
-(d-2)\int\limits_0^1\diff s \int\limits_0^s &\diff t  \detsigtil\updelta_{\bar{b}_1\cdots \bar{b}_{d-3}}^{\bar{a}_1\cdots \bar{a}_{d-3}}k^{\bar{b}_1}_{\bar{a}_1}\left(\frac{1}{2}\tilde{\mathcal{R}}^{\bar{b}_2\bar{b}_3}_{\bar{a}_2\bar{a}_3}-s^2k^{\bar{b}_2}_{\bar{a}_2}k^{\bar{b}_3}_{\bar{a}_3}+\frac{t^2}{\Leff^2}\updelta^{\bar{b}_2}_{\bar{a}_2}\updelta^{\bar{b}_3}_{\bar{a}_3}\right)\times\cdots\\
    &\cdots\times\left(\frac{1}{2}\tilde{\mathcal{R}}^{\bar{b}_{d-4}\bar{b}_{d-3}}_{\bar{a}_{d-4}\bar{a}_{d-3}}-s^2k^{\bar{b}_{d-4}}_{\bar{a}_{d-4}}k^{\bar{b}_{d-3}}_{\bar{a}_{d-3}}+\frac{t^2}{\Leff^2}\updelta^{\bar{b}_{d-4}}_{\bar{a}_{d-4}}\updelta^{\bar{b}_{d-3}}_{\bar{a}_{d-3}}\right) \,, 
\end{split}& \text{even }d \,,
\end{cases}
\end{equation}
where the corresponding terms are defined after eq.\eqref{Bdodd}. 
Having the renormalized EE functional at hand, we evaluate it on certain configurations, i.e. sphere and cylinder, whose universal terms encode significant information for the corresponding CFT.

\section{HEE for spheres in vacuum CFT}

The bulk dual to the vacuum state of a $d$-dimensional CFT is pure AdS$_{d+1}$, whose metric in Poincar\'e coordinates is given by
\begin{equation}\label{eq:AdSdplusonePoincare}
    \diff s^2=\mathcal{G}_{\mu\nu}\diff x^\mu\diff x^\nu=\frac{\Leff^2}{z^2}\left(\diff t^2+\diff z^2+\diff r^2+r^2\diff\Omega_{d-2}^2\right) \,,
\end{equation}
where $\Omega_{d-2}^2$ represents the angular directions of an $\mathbb{S}^{d-2}$ sphere. For ball-shaped entangling regions of radius $R$ in the CFT, the bulk extremal surface is given by the spherical hemisphere of the same radius \cite{Bhattacharyya:2014yga}, whose embedding is described by
\begin{equation}
\Sigma:\left\{t=\text{const.};r^2+z^2=R^2\right\}\label{ExplicitSphere} \,.
\end{equation}
For the following analysis, it is convenient to foliate pure AdS with warped spherical hemispheres. In order to make the extremal surface explicit, the change of coordinates $r=X\sin U$, $z=X\cos U$ is performed. After this change, metric \eqref{eq:AdSdplusonePoincare} reads
\begin{equation}
\diff s^2=\frac{\Leff^2}{X^2\cos^2 U}\left(\diff t^2+\diff X^2 +X^2\diff U^2 + X^2\sin^2U\diff\Omega_{d-2}^2\right) \,.
\end{equation}
In this metric, the hemispheres are the constant $(t,X)$ cod-2 hypersurfaces and the extremal one is located at $X=R$. Also, the non-zero components of the normal vectors to the hypersurfaces read
\begin{equation}
    n_X^{(X)}=n_t^{(t)}=\frac{\Leff}{X\cos U} \,.
\end{equation}
Therefore, the non-zero components of the projected Riemann and Ricci tensors along these directions read
\begin{IEEEeqnarray}{rl}
   R^{(X)(X)}&=R^{\mu\nu}n_\mu^{(X)}n_\nu^{(X)}=-\frac{d}{\Leff^2} \,, \quad R^{(t)(t)}=R^{\mu\nu}n_\mu^{(t)}n_\nu^{(t)}=-\frac{d}{\Leff^2} \,,\\
    R^{(X)(t)(X)(t)}&=R^{\mu\nu\kappa\lambda}n_\mu^{(X)}n_\nu^{(t)}n_\kappa^{(X)}n_\lambda^{(t)}=-\frac{1}{\Leff^2} \,.
\end{IEEEeqnarray}
Regarding the extrinsic curvatures, since the foliation defines a sphere, they identically vanish, i.e.,
\begin{equation}
    {\mathcal{K}^{(X)}}^{\bar{\mu}}_{\bar{\nu}}={\mathcal{K}^{(t)}}^{\bar{\mu}}_{\bar{\nu}}=0 \,.
\end{equation}
In Poincar\'e coordinates, the induced metric reads
\begin{equation}\label{eq:IndSigma}
    \diff s^2_\sigma=\sigma_{\bar{\mu}\bar{\nu}}\diff y^{\bar{\mu}} \diff y^{\bar{\nu}}=\frac{\Leff^2}{z^2}\left[\frac{R^2\diff z^2}{R^2-z^2}+(R^2-z^2)\diff\Omega^2_{d-2}\right] \,,
\end{equation}
which admits a FG-like expansion
\begin{IEEEeqnarray}{rl}\label{eq:IndSigma2}
\diff s^2_\sigma&=\frac{\Leff^2}{z^2}\left[1+\frac{z^2}{R^2}+\frac{z^4}{R^4}+\mathcal{O}\left(z^6\right)\right]\diff z^2+\tilde{\sigma}_{\bar{a}\bar{b}}\diff Y^{\bar{a}}\diff Y^{\bar{b}} \,,\\
\tilde{\sigma}_{\bar{a}\bar{b}}&=\frac{R^2\Leff^2}{z^2}\left(1-\frac{z^2}{R^2}\right)\Omega_{\bar{a}\bar{b}} \,,
\end{IEEEeqnarray}
where $\Omega_{ij}$ is the metric of the $(d-2)$-dimensional sphere. The induced metric $\tilde{\sigma}_{ij}$ is fixed at the regulator $z=\epsilon$, i.e. when $\epsilon\rightarrow0$.

Under the previous considerations, the quantities present in eq.\eqref{SrenQCG} read
\begin{IEEEeqnarray}{rl}
\text{Area}[\Sigma]&=\vol\left(\mathbb{S}^{d-2}\right)R\int\limits_\epsilon^{z_{\max}}\diff z\left(R^2-z^2\right)^{\frac{d-3}{2}}\left(\frac{\Leff}{z}\right)^{d-1} \,, \\
\tilde{R}&=-\frac{(d-1)(d-2)}{\Leff^2} \,, \\
R^{(i)(i)}&=-\frac{2d}{\Leff^2} \,, \\
R^{(i)(j)(i)(j)}&=-\frac{2}{\Leff^2} \,, \\
{\mathcal{K}^{(i)}}^{\bar{\mu}}_{\bar{\mu}}{\mathcal{K}^{(i)}}^{\bar{\nu}}_{\bar{\nu}}={\mathcal{K}^{(i)}}^{\bar{\mu}}_{\bar{\nu}}{\mathcal{K}^{(i)}}^{\bar{\nu}}_{\bar{\mu}}&=0 \,.
\end{IEEEeqnarray}
The cod-2 extrinsic curvatures vanish because the geometry after the foliations describes a sphere.

Therefore, considering all the terms that appear in eq.(\ref{eq:SQCGbare}), we have that the bare HEE $S_{\text{QCG}}$ is given by
\begin{IEEEeqnarray}{rl}
    S_{\text{QCG}}=&\frac{1}{4G_N}\Bigg(\vol\left(\mathbb{S}^{d-2}\right)R\int\limits_\epsilon^{z_{\max}}\diff z\left(R^2-z^2\right)^{\frac{d-3}{2}}\left(\frac{\Leff}{z}\right)^{d-1} \\
    &+\int\limits_\Sigma\diffmone\deth\left\{-\frac{2\alpha}{\Leff^2}\left[(d-1)(d-2)+4d-2\right]-\frac{2d\beta}{\Leff^2}-\gamma\frac{(d-1)(d-2)}{\Leff^2}\right\}\Bigg) \nonumber \,,
\end{IEEEeqnarray}
which can be rearranged, using the definition of $a_d$ in eq.\eqref{eq:ad}, to
\begin{IEEEeqnarray}{rl}
    S_{\text{QCG}}=&\frac{a_d\vol\left(\mathbb{S}^{d-2}\right)}{4G_N}\int\limits_\epsilon^{z_{\max}}\diff z\ R\left(R^2-z^2\right)^{\frac{d-3}{2}}\left(\frac{\Leff}{z}\right)^{d-1} \nonumber \\
    =& a_d \frac{\text{Area}\left[ \Sigma \right]}{4G_N} \,.
\end{IEEEeqnarray}
Also, the HEE Kounterterm $\SKt$ of eq.\eqref{eq:SKt} can be written as
\begin{equation}
\SKt=a_d\frac{c^{\text{EH}}_{d}}{4G_{N}}\left \lfloor{\frac{(d+1)}{2}}\right \rfloor\int\limits_{\partial\Sigma}B_{d-2} \,,
\end{equation}
where $c^{\text{EH}}_{d}=\frac{c_d}{a_d}$ is the Kounterterm coupling for Einstein-AdS gravity. Thus, we have that
\begin{IEEEeqnarray}{rl} \label{QCGuniv}
    S_{\text{QCG}}^{\text{Univ}}=&\frac{a_d}{4G_N} \left[ \text{Area}\left[ \Sigma \right]+ c^{\text{EH}}_{d}\left \lfloor{\frac{(d+1)}{2}}\right \rfloor\int\limits_{\partial\Sigma}B_{d-2} \right] \,,\nonumber \\
    S_{\text{QCG}}^{\text{Univ}}=& \frac{a_d}{4 G_N} \text{Area}_{\text{Univ}}\left[\Sigma\right] \,,
\end{IEEEeqnarray}
such that the universal part of the HEE for ball-shaped entangling regions becomes proportional to the universal part of the area of the minimal surface $\Sigma$.

Finally, using our results of refs. \cite{Anastasiou:2019ldc,Anastasiou:2018rla}, $\text{Area}_{\text{Univ}}$ is given by\footnote{Note the choice of $2R$ as the characteristic scale inside the logarithm of the universal term. This choice allows to absorb the finite term as part of the logarithmically divergent term.}
\begin{equation} \label{areauniv}
\text{Area}_{\text{Univ}}=\begin{dcases*}(-1)^{\frac{d-1}{2}}\frac{2^{d-1}\pi^{\frac{d-1}{2}}\Leff^{d-1}}
{\left( d-1 \right)  !} & \text{if $d$ odd} \,,\\
(-1)^{\frac{d}{2}-1}\frac{2\pi^{\frac{d}{2}-1}\Leff^{d-1}}
{\left( \frac{d}{2}-1 \right)  !} \log\left(\frac{2R}{\epsilon}\right)&  \text{if $d$ even \,,}\end{dcases*}
\end{equation}
where $a_d$ was defined in eq.\eqref{eq:ad}.
The explicit cancellation of the IR divergences in the area functional and the identification of the universal term are given in Appendix \ref{Appendix A}.

\subsection{$C$-function candidates in CFTs dual to QCG}

In order to characterize the properties of CFTs, such as their central charges, it is useful to compute the $C$-function candidates, which are quantities conjectured to decrease along RG flows \cite{Zamolodchikov:1986gt,Cardy:1988cwa,Komargodski:2011vj,Jafferis:2011zi,Klebanov:2011gs,Casini:2012ei}. In the context of AdS/CFT, these quantities can be computed by holographic methods. For instance, in the case of ball-shaped entangling regions, we can read out the $C$-function candidates for both odd and even dimensional CFTs directly from the expression for $S_{\text{QCG}}^{\text{Univ}}$ \cite{Myers:2010xs,Myers:2010tj,Nishioka:2018khk}. In particular, we have
\begin{equation}
S_{\text{QCG}}^{\text{Univ}}=\begin{dcases*}(-1)^{\frac{d-1}{2}}F  & \text{if $d$ odd \,,}\\
(-1)^{\frac{d}{2}-1}4A \log\left(\frac{2R}{\epsilon}\right) &  \text{ if $d$ even\,,}\end{dcases*}\label{LogUniv}
\end{equation}
where
\begin{equation}
F=a_{d}\frac{2^d\pi^{\frac{d-1}{2}}\Leff^{d-1}}
{8G_{N}\left( d-1 \right)  !} \,
\end{equation}
is the $F$-quantity, defined in terms of the partition function of the CFT evaluated on a sphere as $F=\left(-1\right)^{\frac{d-1}{2}}\log\left(Z_{\text{CFT}}[\mathbb{S}^d]\right)$, and
\begin{equation}
    A=a_{d}\frac{\pi^{\frac{d}{2}-1}\Leff^{d-1}}
{8G_{N}\left( \frac{d}{2}-1 \right)  !} \,,
\end{equation}
is the type-A anomaly coefficient. They correspond to quantities that are conjectured to be monotonic along RG flows for odd and even-dimensional CFTs respectively \cite{Myers:2010tj,Myers:2010xs}\footnote{Although the $F$ and $A$ theorems have been proven for arbitrary (unitary) CFTs only in 3 and 4 dimensions respectively, the monotonicity of the $a^{*}$ charge of which they are particular cases is conjectured to apply for arbitrary dimension. Furthermore, the holographic version of the corresponding $C-$theorem has been proven for CFTs dual to Einstein gravity in arbitrary dimension. For the case of CFTs dual to higher curvature gravities, the case of Einsteinian Cubic Gravity was studied in \cite{Bueno:2018xqc}. Therefore, even though the $C-$theorem has not been proven in the generic case, the corresponding $C-$functions for CFTs dual to QCG are expected to be monotonic along RG flows.}. Therefore, the $C$-function candidate for CFTs dual to QCG is proportional to the one of Einstein-AdS gravity, but multiplied by an overall coefficient that depends on the parameters of the theory $a_{d}$. An identical behavior of the universal terms has been found in other higher-curvature theories of gravity such as Einstein cubic gravity, Quasitopological gravity and Lovelock theories, at least, at perturbative level \cite{Bueno:2020uxs}.

The explanation behind this proportionality becomes clear when eqs.\eqref{QCGuniv} and \eqref{areauniv} are considered. Note that the computation of $S_{\text{QCG}}^{\text{Univ}}$ is simplified, for the case of ball-shaped entangling regions, as the resulting functional is proportional to the renormalized area $\text{Area}_{\text{Univ}}$ of the minimal surface in the bulk. This is due to the fact that for spheres, the bare entropy functional becomes proportional to the area of $\Sigma$ what is the RT functional. Therefore, the $C$-function candidates of both QCG and Einstein-AdS gravity correspond to the universal terms of the area of $\Sigma$, given by $\text{Area}_{\text{Univ}}$, up to a factor that depends on the QCG couplings.

The fact that the entropy has to be proportional to the area in the spherical case is universal, and can be inferred directly from the CHM map \cite{Casini:2011kv}. In particular, due to the conformal symmetry of the CFT, the EE of the ball-shaped subregion can be mapped to the thermal entropy of the CFT at a certain temperature that depends on the replica index. This entropy can be computed, using AdS/CFT, as the Wald entropy of a hyperbolic black hole of constant curvature, which is trivially proportional to the area of the black hole horizon.

Note also, that in the expression for the log universal term of eq.\eqref{LogUniv}, one can consider the radius of the sphere $R$ as the characteristic size scale. In which case the logarithmic term can be written as $\log\frac{R}{\epsilon}+\log 2$. This extra $\log2$ appears in even $d$. Because of the robustness of the term in different dimensions, it is suggestive to consider it as coming from a topological term. Indeed, it can be written as $\log\left(\chi\left[\partial \Sigma \right]\right)$, where $\chi\left[\partial \Sigma \right]$ is the Euler characteristic of the entangling surface in the CFT.

In the following section, we consider  cylinder-shaped entangling regions, from which it is possible to compute the type-B anomaly coefficient in four-dimensional CFTs \cite{Bhattacharyya:2014yga,Hung:2011xb}. 

\section{HEE for a cylinder in vacuum CFT}

In order to characterize the type-B anomaly of a CFT, it is useful to consider the log part of the HEE for a cylindrical entangling region. For instance, in the case of AdS$_{5}$/CFT$_{4}$, this universal term is related to $c$ (the type-B anomaly coefficient) according to
\begin{equation}
    S_{\text{EE}}^{\text{Univ}}=-\frac{cH}{2l}\log\frac{l}{\epsilon} \,,
    \label{SUnivC}
\end{equation}
where $l$ is the radius of the cylinder, $H$ is its length along the axis and $\epsilon$ is the usual UV cutoff in the CFT \cite{Bhattacharyya:2014yga,Hung:2011xb}. When computing the EE holographically, by comparing the obtained result with the previous expression, it is possible to identify the $c$ coefficient in terms of the bulk gravity quantities.

We start by considering the metric of pure AdS$_{5}$ written as
\begin{equation}
    \diff s^2=\mathcal{G}_{\mu\nu}\diff x^\mu\diff x^\nu=\frac{\Leff^2}{z^2}\left(\diff t^2+\diff z^2+\diff x_3^2+\diff r^2+r^2\diff\theta^2\right) \,,
\end{equation}
where $\theta$ represents the angular direction of an $\mathbb{S}^{1}$ sphere. For cylindrical entangling regions of radius $l$ in the CFT, with their axis extending infinitely along the $x_3$ direction, the bulk extremal surface, in the near-boundary region, is described by the embedding  
\begin{equation}\label{eq:SigmaCyl}
    \Sigma:\left\{t=\text{const.};r=l\left[1-\frac{z^2}{4l^2}+\mathcal{O}\left(z^4\right)\right]\right\} \,.
\end{equation}
The normal vectors to the hypersurface read
\begin{equation}
    n_\mu^{(X)}=\frac{\Leff}{\sqrt{4l^2+z^2}}\left(0,1,0,\frac{2l}{z},0\right) \,,\quad n^{(t)}_{\mu}=\left(\frac{\Leff}{z},0,0,0,0\right) \,.
\end{equation}
In this case, the projected Riemann and Ricci tensors read
\begin{IEEEeqnarray}{rl}
   R^{(t)(t)}&=R^{tt}n_{t}^{(t)}n_{t}^{(t)}=-\frac{4}{\Leff^2} \,,\\
   R^{(X)(X)}&=R^{rr}n_{r}^{(X)}n_{r}^{(X)}+R^{zz}n_{z}^{(X)}n_{z}^{(X)}=-\frac{4}{\Leff^2} \,,\\
   R^{(X)(t)(X)(t)}&= R^{ztzt}n_{z}^{(X)}n_{t}^{(t)}n_{z}^{(X)}n_{t}^{(t)}+R^{rtrt}n_{r}^{(X)}n_{t}^{(t)}n_{r}^{(X)}n_{t}^{(t)}=-\frac{1}{\Leff^2} \,.
\end{IEEEeqnarray}
The extrinsic curvature along the temporal axis vanishes, i.e., ${\mathcal{K}^{\bar{\mu}}_{\bar{\nu}}}^{(t)}=0$ . However, the foliation in the $z$ coordinates gives a non-zero extrinsic curvature whose components read
\begin{IEEEeqnarray}{rl}
 {\mathcal{K}^{z}_{z}}^{(X)}&=-\frac{z^3}{8 l^3\Leff}+\mathcal{O}\left(z^5\right) \,,\\
  {\mathcal{K}^{x_3}_{x_3}}^{(X)}&=-\frac{z}{2 l\Leff}+\frac{z^3}{16  l^3\Leff}+\mathcal{O}\left(z^5\right) \,,\\
    {\mathcal{K}^{\theta}_{\theta}}^{(X)}&=\frac{z}{2 l\Leff}+\frac{3z^3}{16  l^3\Leff}+\mathcal{O}\left(z^5\right) \,.
\end{IEEEeqnarray}
The induced metric in the cod-2 manifold $\Sigma$ reads
\begin{equation}
    \diff s^2_\sigma=\frac{\Leff^2}{z^2}\left\{\left[1+\frac{z^2}{4l^2}+\mathcal{O}\left(z^4\right)\right]\diff z^2+\diff x_{3}^2+l^2\left[1-\frac{z^2}{4l^2}+\mathcal{O}\left(z^4\right)\right]^2\diff\theta^2\right\} \,.
\end{equation}
As in the spherical entangling region case, this expression admits a FG expansion as well. 

Based on these considerations, the geometric quantities appearing in the EE functional in eq.\eqref{eq:SQCGbare} are given by
\begin{IEEEeqnarray}{rl}
\text{Area}[\Sigma]&=\int\limits_0^{2\pi}\diff\theta\int\limits_0^{H}\diff x_{3}\int\limits_\epsilon^{z_\text{max}}\diff z\deth \,,\\
 \tilde{R}&=-\frac{6}{\Leff^2}-\frac{z^2}{2l^2\Leff^2}+\mathcal{O}\left(z^4\right) \,,\\
R^{(i)(i)}&=-\frac{8}{\Leff^2} \,,\\
R^{(i)(j)(i)(j)}&=-\frac{2}{\Leff^2} \,,\\
{\mathcal{K}^{(i)}}^{a}_{a}{\mathcal{K}^{(i)}}^b_{b}&={\mathcal{O}\left(z^6\right) \,,}\\
{\mathcal{K}^{(i)}}^{a}_{b}{\mathcal{K}^{(i)}}^b_{a}&=\frac{z^2}{2 l^2 \Leff^2}+\mathcal{O}\left(z^4\right) \,.
\end{IEEEeqnarray}
In even-dimensional CFTs, the finite part of the EE is non-universal, and therefore, upon evaluating the integral in the Poincar\'e coordinate of the area functional, the upper limit (at $z_{\text{max}}$) can be neglected. For the lower limit we expand the metric determinant, finding
\begin{equation}
    \detsigma=\frac{l \Leff^3}{z^3}-\frac{\Leff^3}{8 l z}+\mathcal{O}\left(z\right) \,.
\end{equation}
Now, plugging all these results into the functional, we obtain
\begin{equation}\label{eq:Sbarecyl}
    S_{\text{QCG}}=\frac{\pi H\Leff^3}{4l G_N}\left(a_4\frac{l^2}{\epsilon^2}-\frac{b_4}{4}\log\frac{l}{\epsilon}\right)+\mathcal{O}(1) \,,
\end{equation}
where we define the coefficient
\begin{equation}
b_4=1-\frac{4}{\Leff^2}\left(10\alpha+2\beta+\gamma\right) \,.\label{b_4}
\end{equation}
This factor differs from $a_4$ defined in eq.\eqref{eq:ad}. 

Finally, we check that the boundary term cancels the power law term in eq.\eqref{eq:Sbarecyl}. The induced metric at the boundary $z=\epsilon$ reads
\begin{equation}
    \sqrt{\tilde{\sigma}}=\frac{l\Leff^2}{\epsilon^2}-\frac{\Leff^2}{4l}+\mathcal{O}\left(\epsilon\right) \,,
\end{equation}
yielding
\begin{equation}
\SKt=-\frac{a_4\pi  H l \Leff^3}{4 G_{N} \epsilon ^2} \,.
\end{equation}
As we can see, the power law divergence in eq.\eqref{eq:Sbarecyl} is indeed cancelled by the Kounterterm.

Thus, up to a non-universal finite part, one has that for the cylinder entangling region in $d=4$, the universal part of the HEE is given by
\begin{eqnarray}
    S_{\text{QCG}}^{\text{Univ}}=-b_{4}\frac{\pi H \Leff^3}{16 l G_N}\log \frac{l}{\epsilon} \,.
\end{eqnarray}
Finally, comparing this expression with that of eq.(\ref{SUnivC}), we have that
\begin{eqnarray}\label{c_Coeff}
    c=b_{4}\frac{\pi \Leff^3}{8 G_N} \,.
\end{eqnarray}
In the QCG case, it is evident from our results that the $A$ and $c$ central charges are different. However, for Einstein-AdS ($b_{4}=a_{4}=1$), they coincide.

\subsection{Extremal surface for the cylinder}

In refs.\cite{Bhattacharyya:2014yga,Hung:2011xb}, it was shown that the hypersurface \eqref{eq:SigmaCyl} extremizes the HEE functional for Gauss-Bonnet theory. In QCG, higher-order terms appear in the entanglement entropy. However, the same embedding function yields the extremal surface at order $\mathcal{O}\left(z^3\right)$. In order to see this, consider an arbitrary surface parametrized with $r=r(z)$. In $d=4$, the HEE functional for QCG reads
\begin{IEEEeqnarray}{cl}\label{eq:Cylfunc}
 S_{\text{QCG}}=&\frac{\pi H\Leff}{4G_{N}}\int\limits_\epsilon^{z_\text{max}}\diff z\frac{1}{z^3 r
\left(r'^2+1\right)^{5/2}}\Big\{2 \Leff^2 r^2 \left(r'^2+1\right)^3\nonumber\\
&+16\alpha r(r'^2+1)\left[z(z+2rr')r''-(r'^2+1)(r(5+3r'^2)-z r')\right]
\nonumber\\
   &-\beta  \left[\left(r \left(r'^3+r'+z
   r''\right)+z \left(r'^2+1\right)\right)^2+16 r^2 \left(r'^2+1\right)^3\right]\nonumber\\
   &-   4
   \gamma  r \left(r'^2+1\right) \left[2 z \left(2 r r'+z\right) r''+2 \left(3
   r-2 z r'\right) \left(r'^2+1\right)\right]\Big\} \,.
\end{IEEEeqnarray}
This expression is obtained once the terms given by
\begin{IEEEeqnarray}{rl}
\text{Area}[\Sigma]&=\int\limits_0^{2\pi}\diff\theta\int\limits_0^{H}\diff u\int\limits_\epsilon^{z_\text{max}}\diff z\detsigma=2\pi H \Leff^3\int\limits_\epsilon^{z_\text{max}}\diff z \frac{r\sqrt{1+r'^2}}{z^3} \,,\\
\tilde{R}&=-\frac{2 z \left(2 r r'+z\right) r''+2 \left(3 r-2 z r'\right)
   \left(r'^2+1\right)}{\Leff^2 r \left(r'^2+1\right)^2} \,,\\
R^{(i)(i)}&=-\frac{8}{\Leff^2} \,,\\
R^{(i)(j)(i)(j)}&=-\frac{2}{\Leff^2}\,,\\
{\mathcal{K}^{(i)}}^{\bar{\mu}}_{\bar{\mu}}{\mathcal{K}^{(i)}}^{\bar{\nu}}_{\bar{\nu}}&=\frac{\left[\left(r r'+z\right) \left(r'^2+1\right)+z r r''\right]^2}{\Leff^2 r^2
   \left(r'^2+1\right)^3} \,,\\
{\mathcal{K}^{(i)}}^{\bar{\mu}}_{\bar{\nu}}{\mathcal{K}^{(i)}}^{\bar{\nu}}_{\bar{\mu}}&=\frac{\left[(z+r r')^2+r^2r'^2\right](1+r'^2)^2+r^2(r'+r'^3-zr'')}{\Leff^2r^2(1+r'^2)^3} \,,
\end{IEEEeqnarray}
are plugged into eq.\eqref{eq:SQCGbare}. The resulting functional \eqref{eq:Cylfunc} constitutes a Lagrangian $\mathcal{L}=\mathcal{L}(z,r,r',r'')$ that contains second order derivatives of the dynamical function $r(z)$. Because of this, the Euler-Lagrange equation needed to find the extremal surface reads
\begin{equation}
    \frac{\partial \mathcal{L}}{\partial r}-\frac{\diff}{\diff z}\frac{\partial \mathcal{L}}{\partial r'}+\frac{\diff^2}{\diff z^2}\frac{\partial \mathcal{L}}{\partial r''}=0 \,.
    \label{EL_HEE}
\end{equation}
From this expression, an EOM containing fourth-derivative terms in the function $r(z)$ is found. However, the ansatz
\begin{equation}
    r=l\left[1-\frac{z^2}{4l^2}+\mathcal{O}\left(z^4\right)\right] \,,
    \label{cylindemb}
\end{equation}
from ref.\cite{Hung:2011xb} is verified to satisfy the EOM up to order $\mathcal{O}\left(z^3\right)$. Because of this, the extremal surface for QCG coincides with that for Gauss-Bonnet gravity in this perturbative regime. This is an expected result due to the universality of the second term in the asymptotic expansion of the embedding function, what is linked to the universality of the $g_{(2) ij}$ coefficient in the FG expansion in terms of the Schouten tensor of $g_{(0)}$, as discussed in ref.\cite{Schwimmer:2008yh}.

The case for the cylindrical entangling region in $d=4$ is interesting as it isolates the contribution from the type-B anomaly in the universal part. The same should be the case for higher-dimensional cylinders, as the coefficient obtained should represent a linear combination of the couplings of different conformal invariants.\footnote{Although little is known about conformal invariants beyond eight dimensions, one may think this computation would provide information on the part of these invariants which is polynomial in the Weyl tensor.} However, the embedding function is not known in the higher-dimensional case as it would require knowledge of the subleading terms in the expansion of eq.\eqref{cylindemb}, which are not universal. Furthermore, the Kounterterm renormalization procedure has limitations regarding the types of entangling surfaces that it can accommodate, for dual bulk manifolds of dimension greater than 5. In particular, it requires the dimensional continuation of cod-2 conformal invariants at the entangling surface to vanish. For example, to the next-to-leading order, the method only works for surfaces such that

\begin{equation}
\mathcal{W}_{\overline{a} \overline{b}}^{\overline{a} \overline{b}}-\hat{k}_{\overline{a} \overline{b}}^{\left(\overline{i}\right)} \hat{k}^{\overline{a} \overline{b}}_{\left(\overline{i}\right)}   =0 \,,
\label{Kountershapelimit}
\end{equation}
as shown in ref.\cite{Anastasiou:2019ldc}. Here, $\mathcal{W}_{\overline{a} \overline{b}}^{\overline{a} \overline{b}}$ is the Weyl tensor of the conformal boundary projected along the entangling surface directions and $\hat{k}_{\overline{a} \overline{b}}$ is the traceless extrinsic curvature of $\partial \Sigma$ along the orthogonal directions. Eq.\eqref{Kountershapelimit} is trivially satisfied for spheres, but not for cylinders or arbitrary shapes.\footnote{In bulk dimensions up to 5, the Kounterterm procedure works for arbitrary entangling regions. Furthermore, as proven in \cite{Araya:2021atx}, the procedure correctly renormalizes actions for gravity theories of arbitrary order in the Riemannian curvature, and therefore, it is expected to work for renormalizing HEE for CFTs dual to said theories as well.}

\section{Discussion}

The results for HEE for CFTs dual to QCG presented here come as the natural blend between the Kounterterm method applied to this gravity theory \cite{Giribet:2018hck,Giribet:2020aks} and a remarkable feature of the boundary term $B_{d}$ when evaluated in spacetimes with a conical defect, eq.(\ref{Kounterevenorbifold}) \cite{Anastasiou:2019ldc}.  In this respect, we have recovered the universal part of the HEE found in the literature  regarding the computation of the $C$-function candidates \cite{Bueno:2020uxs}. This function captures essential properties of CFTs, which are given by the type-A anomaly coefficient in the case of even $d$ and by the generalized $F$ quantity (or $a^{*}$ charge) for odd $d$ \cite{Myers:2010xs,Nishioka:2018khk,Imbimbo:1999bj}.

The above calculation requires a ball-shaped entangling region in the CFT, where for the case of pure AdS (dual to the vacuum of the CFT), the embedding of the extremal surface \eqref{ExplicitSphere} for the QCG HEE functional \eqref{eq:SQCGbare} is explicitly given. Then, the FG-like expansion of all the terms involved in the functional can be obtained. In both even and odd boundary dimension $d$, it can be seen that the  $C$-function candidate derived is proportional to the one for Einstein-AdS gravity, but with an overall coupling-dependent factor $a_{d}$, whose form is given in eq.\eqref{eq:ad}.

\sloppy 
We have also obtained the type-B anomaly coefficient $c$ in the case of four-dimensional CFTs. In order to perform this computation, we have considered a cylindrical entangling region in the CFT, and the near-boundary expansion (up to cubic order in the Poincar\'e coordinate) of the embedding for the minimal surface. In this situation, we have derived the corresponding Euler-Lagrange equation for the embedding function $r(z)$, by taking variations of the entropy functional of eq.\eqref{eq:Cylfunc}. We have verified that the same embedding function considered for Einstein-AdS gravity is also a solution of the extremization equation in the QCG case, in $d=4$, and up to cubic order. The condition for the minimal surface obtained in this way contains higher-order derivative terms in the dynamical variable (akin to the \emph{acceleration}). Thus, extra boundary conditions are required beyond setting the border of the surface to coincide with the entangling region. When the surface is extremal (i.e., the intersection of the surface with the conformal boundary is orthogonal), the boundary problem is completely fixed. This is the case of the cylinder in $d=4$, due to the fact that the entropy functional becomes proportional to the area up to cubic order. In the result for the $c$ coefficient we also find agreement with the literature \cite{Bueno:2020uxs}. It is evident from the  expression obtained in eq.\eqref{c_Coeff}, that the type-B anomaly coefficient can be written as the one for Einstein-AdS gravity, but multiplied with a factor $b_{4}$, given in eq.\eqref{b_4}, which incorporates the information on the couplings of QCG theory. The fact that $b_{4}$ is different from the $a_{4}$ of eq.\eqref{eq:ad} allows for different central charges in four-dimensional CFTs dual to QCG, unlike the Einstein-AdS case.

All in all, for both even and odd dimensional CFT cases, the Kounterterm procedure allows to isolate the universal part of the HEE of the dual gravity theory. For bulk dimensions lower than 6, the Kounterterm procedure works on entangling regions of arbitrary shape. Also, despite its limitations on the type of entangling regions that can be renormalized in higher dimensions (as discussed in the previous section), the Kounterterms prescription is the only method available so far for renormalizing HEE in higher curvature gravity theories.\footnote{The alternative renormalization procedure of \cite{Taylor:2016aoi}, based on Holographic Renormalization \cite{deHaro:2000vlm}, was only applied for Einstein and Einstein-Gauss-Bonnet gravity theories.} 

In isolating the universal part of the HEE, we have been able to express it as a covariant functional which is given by the standard HEE functional plus an extrinsic counterterm in cod-3 \eqref{eq:SKt}. In the particular case of spherical entangling regions in pure AdS (vacuum CFTs), the renormalized HEE functional becomes proportional to the renormalized area \eqref{QCGuniv}, which is logarithmically divergent for even-$d$ and finite for odd-$d$. 

For a cylindrical entangling region in $d>4$, the renormalized entropy functional is no longer proportional to the renormalized cod-2 volume. However, in $d=4$, the expressions coincide -up to the normalizable order-, albeit with a different proportionality constant than for the spherical case.

We point out that the Kounterterm procedure is a nonperturbative method, in the sense that nowhere it is assumed that the couplings of the quadratic terms are small. In other words, the prescription does not rely on the linearization of the EOM such that the theory behaves like Einstein gravity with a modified Newton's constant. In particular,  Kounterterms give a consistent renormalization prescription even at degenerate points, such as at the single-vacuum point of Einstein-Gauss-Bonnet theory (Chern-Simons AdS gravity in five dimensions). Thus, the obtained form of the holographic central charges is robust, and would hold in degenerate points of the parametric space as well (in the case of Lovelock, see ref.\cite{Arenas-Henriquez:2019rph}). Although we have not developed this point any further in this work, it would certainly be interesting to explore it in the near future.

\acknowledgments

We thank Andr\'es Argando\~na for useful comments. The work of GA and RO was funded in part by FONDECYT grants No. 3190314 \textit{Holographic Complexity from Anti-de Sitter gravity} and No. 1170765 \textit{Boundary dynamics in anti-de Sitter gravity and gauge/gravity duality}. The work of JM is funded by the Agencia Nacional de Investigaci\'on y Desarrollo (ANID) Scholarship No. 21190234 and by Pontificia Universidad Cat\'olica de Valpara\'iso. JM is also grateful to the QMAP faculty for their hospitality. The work of IJA is funded by ANID, REC Convocatoria Nacional Subvenci\'on a Instalaci\'on en la Academia Convocatoria A\~no 2020, Folio PAI77200097. The work of DRB is funded by Becas Chile (ANID) Scholarship No. 72200301.

\appendix

\section{Notation and conventions}

Here, we present the conventions used throughout the paper in the Table \ref{tab:notation}. In the first column, we provide a list of objects defined on the different manifolds presented in the first line.
\begin{table}[H]
    \centering
    \begin{tabular}{|l|c|c|c|c|}
     \cline{2-5}
    \multicolumn{1}{c|}{}
 & $\mathcal{M}$ &  $\partial\mathcal{M}$ & $\Sigma$ & $\partial \Sigma$\\ \hline\hline
        Indices & $\mu,\nu,\kappa,\lambda$ &$a,b,c,d$& $\bar{\mu},\bar{\nu},\bar{\kappa},\bar{\lambda}$ & $\bar{a},\bar{b},\bar{c},\bar{d}$ \\
    Coordinates & $x^\mu$ &$X^a$& $y^{\bar{\mu}}$ & $Y^{\bar{a}}$\\
    Metric & $\mathcal{G}_{\mu\nu}$ & $h_{ab}$ & $\sigma_{\bar{\mu}\bar{\nu}}$& $\tilde{\sigma}_{\bar{a}\bar{b}}$ \\
    Riemann tensor & $R_{\mu\nu}^{\kappa\lambda}$ & $\mathcal{R}_{ab}^{cd}$ & $\tilde{R}_{\bar{\mu}\bar{\nu}}^{\bar{\kappa}\bar{\lambda}}$ & $\tilde{\mathcal{R}}_{\bar{a}\bar{b}}^{\bar{c}\bar{d}}$\\
    Extrinsic curvature & & $K_{ab}$ &$\mathcal{K}_{\bar{\mu}\bar{\nu}}$& $k_{\bar{a}\bar{b}}$\\ \hline
    \end{tabular}
    \caption{Notation and conventions}\label{tab:notation}
\end{table}

\section{Divergence cancellation for spherical entangling regions }\label{Appendix A}

In this section, we present explicit computations and show the cancellation of divergences in the Kounterterms scheme for spherical entangling regions.

\subsection{Three-dimensional case}

For three dimensions, the universal part of the EE \eqref{SrenQCG} reduces to
\begin{IEEEeqnarray}{rl}\label{sd1}
S_{\text{QCG}}^{\text{Univ}}=\frac{\pi R}{2G_N}\int\limits_{\epsilon}^{z_{\max}}\frac{\diff z}{z^2}\left(\Leff^2-24\alpha-6\beta    \right)+\frac{c_3}{2G_N}\int\limits_{\partial \Sigma}B_1 \,,
\end{IEEEeqnarray}
where the auxiliary function $c_3$ and the boundary term $B_1$, defined in eqs.\eqref{eq:cd} and \eqref{eq:Bdminus2} respectively read
\begin{IEEEeqnarray}{rl}
c_3&=\frac{1}{4}\left(\Leff^2-24\alpha-6\beta\right) \,,\\
B_1&=-2\sqrt{\tilde{\sigma}}\tr k\ \diff \Omega_1=-\frac{2}{\epsilon}R\left[1+\mathcal{O}\left(\epsilon^2\right)\right]\diff \Omega_1 \,.
\end{IEEEeqnarray}
On the other hand, the determinant of the metric $\tilde{\sigma}$ is given by
\begin{equation}\label{eq:dettildeh}
    \sqrt{\tilde{\sigma}}=\frac{R\Leff}{\epsilon}\sqrt{1-\frac{\epsilon^2}{R^2}}=\Leff\left[\frac{R}{\epsilon}-\frac{\epsilon}{2R}-\frac{\epsilon^3}{R^3}+\mathcal{O}\left(\epsilon^4\right)\right] \,.
\end{equation}
Also, $\tr k$ is the trace of the extrinsic curvature $k_{\bar{a}\bar{b}}=-\frac{1}{2\sqrt{\sigma_{zz}}}\partial_z\tilde{\sigma}_{\bar{a}\bar{b}}$ of the FG-like expansion \eqref{eq:IndSigma2}, which reads
\begin{equation}
    k_{\bar{a}\bar{b}}=\Leff\left[\frac{R^2}{\epsilon^2}-\frac{1}{2}-\frac{\epsilon^2}{8R^2}+\mathcal{O}\left(\epsilon^4\right)\right]\Omega_{\bar{a}\bar{b}} \,.
\end{equation}
Since the inverse metric reads,
\begin{equation}
    \tilde{\sigma}^{\bar{a}\bar{b}}=\frac{1}{\Leff^2}\left[\frac{\epsilon^2}{R^2}+\mathcal{O}\left(\epsilon^4\right)\right]\Omega^{\bar{a}\bar{b}} \,,
\end{equation}
then, the expansion of the trace yields
\begin{equation}
    \tr k=\tilde{\sigma}^{\bar{a}\bar{b}}k_{\bar{a}\bar{b}}=\frac{1}{\Leff}\left[1+\frac{\epsilon^2}{2R^2}+\mathcal{O}\left(\epsilon^4\right)\right]\,.
\end{equation}
In consequence, according to eq.\eqref{eq:SKt} and up to leading order, the Kounterterm in this case reads
\begin{equation}\label{eq:Sd3}
    \SKt=-\frac{\pi R}{2G_N \epsilon}
    \left[\Leff^2-24\alpha-6\beta\right]+\mathcal{O}(\epsilon) \,.
\end{equation}
Thus, upon performing the integral in eq.\eqref{sd1} on the extremal surface, along the Poincaré coordinate from $z=\epsilon$ to $z=R$, one gets
\begin{IEEEeqnarray}{rl}
S^{\text{Univ}}_{\text{QCG}}=S_{\text{finite}}+\left(\Leff^2-24\alpha-6\beta\right)\left[\frac{\pi R}{2G_N \epsilon}-\frac{\pi R}{2G_N \epsilon}\right]+
    \mathcal{O}(\epsilon) \,,
\end{IEEEeqnarray}
where $S_{\text{finite}}=-\frac{\pi}{2G}\left(\Leff^2-24\alpha-6\beta\right)$ is the universal finite part. Thus, it becomes manifest that upon taking the $\epsilon \rightarrow 0$ limit, $S^{\text{Univ}}_{\text{QCG}}$ recovers the universal finite part and thus the HEE is renormalized correctly. 

\subsection{Four-dimensional case}

In the four-dimensional case, from eq.\eqref{eq:Bdminus2} it can be seen that the corresponding boundary term is given by
\begin{equation}
    B_{2}=-2\Leff\frac{\sqrt{R^2-\epsilon^2}}{\epsilon^2} \,,
\end{equation}
and hence, the Kounterterm expanded around $\epsilon=0$ reads
\begin{equation}
    \SKt=-\frac{a_4\pi\Leff^{3}}{2G_N}\left(\frac{R^2}{\epsilon^2}-\frac{1}{2}\right) \,.
\end{equation}
On the other hand, the bare EE is given by
\begin{IEEEeqnarray}{rl}
    S_{\text{QCG}}=&\frac{a_4\pi}{G_N}\int\limits_\epsilon^{z_{\max}}\diff z\ R\left(R^2-z^2\right)^{\frac{1}{2}}\left(\frac{\Leff}{z}\right)^{3}\\
    =&\frac{a_4\pi\Leff^3}{G_N}\left[\coth^{-1}\left(\frac{R}{\sqrt{R^2-z_\text{max}^2}}\right)-\frac{R}{z_\text{max}^2}\sqrt{R^2-z_\text{max}^2}\right]\nonumber\\
    &+\frac{a_4\pi\Leff^3}{2G_N}\left[\frac{R^2}{\epsilon^2}-\log\frac{R}{\epsilon}-\frac{1}{2}\left(1+\log 4\right)\right] \,.\label{eq:S4d}
\end{IEEEeqnarray}
Since the first line of \eqref{eq:S4d} vanishes in the limit $z_\text{max}\rightarrow R$, then,
after adding the Kounterterm, the renormalized EE reads
\begin{equation}\label{eq:RenS4}
    S^{\text{Univ}}_{\text{QCG}}=-\frac{a_4 \pi \Leff^3}{2G_N}\log\frac{2R}{\epsilon} \,.
\end{equation}
Notice that the Kounterterm isolates the logarithmic divergence, whose coefficient is universal and related to the type-A charge of the CFT. A part of the finite term is cancelled and there is a piece left that is reabsorbed in the logarithmic divergence.

\subsection{Bare HEE in the arbitrary dimensional case}

For arbitrary dimensions, evaluating the quantities present in the EE expression using the metric \eqref{eq:AdSdplusonePoincare} yields
\begin{IEEEeqnarray}{rl}
    S_{\text{QCG}}=&\frac{1}{4G_N}\Bigg(\vol\left(\mathbb{S}^{d-2}\right)R\int\limits_\epsilon^{z_{\max}}\diff z\left(R^2-z^2\right)^{\frac{d-3}{2}}\left(\frac{\Leff}{z}\right)^{d-1}\\
    &+\int\limits_\Sigma\diffmone\deth\left\{-\frac{2\alpha}{\Leff^2}\left[(d-1)(d-2)+4d-2\right]-\frac{2d\beta}{\Leff^2}-\gamma\frac{(d-1)(d-2)}{\Leff^2}\right\}\Bigg) \,,\nonumber
\end{IEEEeqnarray}
which can be rearranged, using definition of $a_d$ in eq.\eqref{eq:ad}, to
\begin{equation}
    S_{\text{QCG}}=\frac{a_d\vol\left(\mathbb{S}^{d-2}\right)}{4G_N}\int\limits_\epsilon^{z_{\max}}\diff z\ R\left(R^2-z^2\right)^{\frac{d-3}{2}}\left(\frac{\Leff}{z}\right)^{d-1} \,.
\end{equation}
After computing the integral and expanding around $\epsilon=0$, the EE yields
\begin{equation}\label{eq:bareSdodd}
        S_{\text{QCG}}=S_{\text{QCG}}^{\text{Univ}}+\frac{a_d\vol\left(\mathbb{S}^{d-2}\right)\Leff^{d-1}}{4G_N}\left[\frac{1}{d-2}\frac{R^{d-2}}{\epsilon^{d-2}}-\frac{d-3}{2(d-4)}\frac{R^{d-4}}{\epsilon^{d-4}}+\mathcal{O}\left(\epsilon^{-(d-6)}\right)\right] \,.
\end{equation}
In the previous expression, the universal term is given by
\begin{equation}
S_{\text{QCG}}^{\text{Univ}}=\begin{dcases*}(-1)^{\frac{d-1}{2}}\frac{a_d(4\pi)^{\frac{d-1}{2}}\left(\frac{d-1}{2}\right)!\Leff^{d-1}}{4G(d-1)!} & \text{if $d$ odd} \,,\\
(-1)^{\frac{d}{2}-1}\frac{a_d\pi^{\frac{d}{2}-1}\Leff^{d-1}}{2G_N(\frac{d}{2}-1)!}\log\frac{2R}{\epsilon}&  \text{if $d$ even \,,}\end{dcases*}
\end{equation}
where $z_\text{max}$ is set to the radius $R$ and the volume of the sphere is given by 
\begin{equation}
\vol(\mathbb{S}^{d-2})=\frac{2^d\pi^{\frac{d}{2}-1}(\frac{d}{2})!}{d(d-2)} \,.
\end{equation}

\subsection{Cancellation of divergences for odd dimensions}

Let us now focus on the Kounterterm. In the odd dimensional case, the boundary form can be read, as before, from eq.\eqref{eq:Bdminus2}.
Since we are considering a spherical entangling region, the extrinsic curvature and the Riemann tensor read, respectively,
\begin{equation}\label{eq:sphECRT}
    k^{\bar{a}}_{\bar{b}}=\frac{1}{\Leff}\frac{R}{\sqrt{R^2-\epsilon^2}}\updelta^{\bar{a}}_{\bar{b}} \,,\quad \tilde{\mathcal{R}}^{\bar{a}_1\bar{a}_2}_{\bar{b}_1\bar{b}_2}=\frac{1}{\Leff^2}\frac{\epsilon^2}{R^2-\epsilon^2}\updelta^{\bar{a}_1\bar{a}_2}_{\bar{b}_1\bar{b}_2}\,.
\end{equation}
Plugging these values into the boundary form, we find
\begin{align}
    B_{d-2} & =-(d-1)\int\limits_0^1\detsigtil\updelta^{\bar{a}_1\cdots \bar{a}_{d-2}}_{\bar{b}_1\cdots \bar{b}_{d-2}}\updelta^{\bar{b}_1}_{\bar{a}_1}\frac{1}{\Leff}\frac{R}{\sqrt{R^2-\epsilon^2}}\left(\frac{1}{2\Leff^2}\frac{\epsilon^2}{R^2-\epsilon^2}\updelta^{\bar{b}_2\bar{b}_3}_{\bar{a}_2\bar{a}_3}-s^2\frac{1}{\Leff^2}\frac{R^2}{R^2-\epsilon^2}\updelta^{\bar{b}_2}_{\bar{a}_2}\updelta^{\bar{b}_3}_{\bar{a}_3}\right)\times\cdots \nonumber\\
    & \cdots\times\left(\frac{1}{2\Leff^2}\frac{\epsilon^2}{R^2-\epsilon^2}\updelta^{\bar{b}_{d-3}\bar{b}_{d-2}}_{\bar{a}_{d-3}\bar{a}_{d-2}}-s^2\frac{1}{\Leff^2}\frac{R^2}{R^2-\epsilon^2}\updelta^{\bar{b}_{d-3}}_{\bar{a}_{d-3}}\updelta^{\bar{b}_{d-2}}_{\bar{a}_{d-2}}\right) \,.
\end{align}
In this expression, we make use the relations
\begin{IEEEeqnarray}{rl}
&\updelta^{\bar{a}_1\cdots \bar{a}_{d-3}}_{\bar{b}_1\cdots \bar{b}_{d-3}}\updelta^{\bar{b}_2\bar{b}_3}_{\bar{a}_2\bar{a}_3}=2\updelta^{\bar{a}_1\cdots \bar{a}_{d-3}}_{\bar{b}_1\cdots {d-3}}\updelta^{\bar{b}_2}_{\bar{a}_2}\updelta^{\bar{b}_3}_{\bar{a}_3} \,,\nonumber\\
\updelta_{\bar{b}_1\cdots \bar{b}_{d-2}}^{\bar{a}_1\cdots \bar{b}_{d-2}}\updelta^{\bar{b}_{d-2}}_{\bar{a}_{d-2}}&=\updelta_{\bar{b}_1\cdots \bar{b}_{d-3}}^{\bar{a}_1\cdots \bar{a}_{d-3}} \,,\quad \updelta_{\bar{b}_1\cdots \bar{b}_{d-3}}^{\bar{a}_1\cdots \bar{a}_{d-3}}\updelta^{\bar{b}_1}_{\bar{a}_1}\cdots\updelta^{\bar{b}_{d-3}}_{\bar{a}_{d-3}}=(d-2)!\ \,.
\end{IEEEeqnarray}
Besides, writing explicitly the determinant $\sqrt{\tilde{\sigma}}$ given in eq.\eqref{eq:dettildeh}, the boundary form reads
\begin{equation}
     B_{d-2}=-(d-1)!\int\limits_0^1\diff s\left[\frac{R^2}{\epsilon^2}\left(1-\frac{\epsilon^2}{R^2}\right)\right]^{\frac{d-2}{2}}\frac{R}{\sqrt{R^2-\epsilon^2}}\left(\frac{\epsilon^2}{R^2-\epsilon^2}-s^2\frac{R^2}{R^2-\epsilon^2}\right)^{\frac{d-3}{2}} \,.
     \end{equation}
     This expression can be expanded around $\epsilon=0$ as
     \begin{equation}
         B_{d-2}=-(d-1)!\int\limits_0^1\diff s\left[(-1)^{\frac{d+1}{2}}s^{d-3}\frac{R^{d-2}}{\epsilon^{d-2}}-(-1)^{\frac{d+1}{2}}s^{d-5}\frac{d-3}{2}\frac{R^{d-4}}{\epsilon^{d-4}}+\mathcal{O}\left(\epsilon^{-(d-6)}\right)\right] \,,
     \end{equation}
which after the integration reads
\begin{equation}
    B_{d-2}=-(d-1)!\left[(-1)^{\frac{d+1}{2}}\frac{1}{d-2}\frac{R^{d-2}}{\epsilon^{d-2}}-(-1)^{\frac{d+1}{2}}\frac{d-3}{2(d-4)}\frac{R^{d-4}}{\epsilon^{d-4}}+\mathcal{O}\left(\epsilon^{-(d-6)}\right)\right] \,.
\end{equation}
Therefore the Kounterterm $\SKt$, after using the definition $c_d$ in the odd dimensional case from eq.\eqref{eq:ad}, becomes
\begin{equation}
   \SKt=-\frac{a_d\vol\left(\mathbb{S}^{d-2}\right)\Leff^{d-1}}{4G_N}\left[\frac{1}{d-2}\frac{R^{d-2}}{\epsilon^{d-2}}-\frac{d-3}{2(d-4)}\frac{R^{d-4}}{\epsilon^{d-4}}+\mathcal{O}\left(\epsilon^{-(d-6)}\right)\right] \,.
\end{equation}
Therefore, adding the previous expression to the bare EE found in eq.\eqref{eq:bareSdodd}, the renormalized EE reads
\begin{equation}
    S^{\text{Univ}}_{\text{QCG}}=S_{\text{finite}} \,.
\end{equation}

\subsection{Cancellation of divergences in even dimensions}

Proceeding as in the odd-dimensional case, the extrinsic curvature and the Riemann tensor, given in eq.\eqref{eq:sphECRT}, are used along with the identities
\begin{IEEEeqnarray}{rl}
\updelta_{\bar{b}_1\cdots \bar{b}_{d-3}}^{\bar{a}_1\cdots \bar{a}_{d-3}}\updelta_{\bar{b}_2\bar{b}_3}^{\bar{a}_2\bar{a}_3}=&2\updelta_{\bar{a}_1\cdots \bar{a}_{d-3}}^{\bar{b}_1\cdots \bar{b}_{d-3}}\updelta^{\bar{b}_2}_{\bar{a}_2}\updelta^{\bar{b}_3}_{\bar{a}_3} \,,\quad 
\updelta_{\bar{b}_1\cdots \bar{b}_{d-3}}^{\bar{a}_1\cdots \bar{a}_{d-3}}\updelta^{\bar{b}_{d-3}}_{\bar{a}_{d-3}}=2\updelta_{\bar{b}_1\cdots \bar{b}_{d-4}}^{\bar{a}_1\cdots \bar{a}_{d-4}} \,,\nonumber\\
&\updelta^{\bar{a}_1\cdots \bar{a}_{d-4}}_{\bar{b}_1\cdots \bar{b}_{d-4}}\updelta^{\bar{b}_1}_{\bar{a}_1}\cdots\updelta^{\bar{b}_{d-4}}_{\bar{a}_{d-4}}=\frac{(d-2)!}{2} \,.
\end{IEEEeqnarray}
After implementing all these relations and substituting $\detsigtil$ the boundary form reads
\begin{equation}
    B_{d-2}=-(d-2)^2(d-3)!\Leff\int\limits_0^1\diff s\int\limits_0^s\diff t\ b(t,s,\epsilon) \,,
\end{equation}
where the  the introduced function $b(t,s)$ reads
\begin{equation}
    b(t,s,\epsilon)=\frac{R^{d-2}}{\epsilon^{d-2}}\left(1-\frac{\epsilon^2}{R^2}\right)^{d-2}\frac{R}{\sqrt{R^2-\epsilon^2}}\left(\frac{\epsilon^2}{R^2-\epsilon^2}-s^2\frac{R^2}{R^2-\epsilon^2}+t^2\right)^{\frac{d}{2}-2}\,.
\end{equation}
If the function $b(t,s,\epsilon)$ is expanded around $\epsilon=0$, then
\begin{equation}
    b(t,s,\epsilon)=\left(t^2-s^2\right)^{\frac{d}{2}-2}\frac{R^{d-2}}{\epsilon^{d-2}}+\frac{1}{2}\left(t^2-s^2\right)^{\frac{d}{2}-3}\left[d-4+s^2-(d-3)t^2\right]\frac{R^{d-4}}{\epsilon^{d-4}}+\mathcal{O}\left(\epsilon^{-(d-6)}\right) \,.
\end{equation}
In this expression, the convergence condition demands that the order $\mathcal{O}\left(\epsilon^{-(d-2)}\right)$ term appears for $d>2$. Likewise, the order $\mathcal{O}\left(\epsilon^{-(d-4)}\right)$ term appears when $d>4$ and successively. Now, computing the integrals
\begin{IEEEeqnarray}{rl}
\int\limits_0^1\diff s\int\limits_0^s\diff t\left(t^2-s^2\right)^{\frac{d}{2}-2}&=\frac{(-1)^{\frac{d}{2}}2^{d-2}\left(\frac{d}{2}-1\right)!^2}{(d-2)^2(d-2)!} \,,\\
\int\limits_0^1\diff s\int\limits_0^s\diff t\left(t^2-s^2\right)^{\frac{d}{2}-3}\left[d-4+s^2-(d-3)t^2\right]&=-\frac{(-1)^{\frac{d}{2}}2^{d-2}(d-3)\left(\frac{d}{2}-1\right)!^2}{(d-2)(d-4)(d-2)!}\,,\nonumber\\
\end{IEEEeqnarray}
the boundary term reduces to
\begin{equation}
        B_{d-2}=(-1)^{\frac{d}{2}+1}2^{d-2}\Leff\left[\left(\frac{d}{2}-1\right)!\right]^2\left[\frac{1}{d-2}\frac{R^{d-2}}{\epsilon^{d-2}}-\frac{d-3}{2(d-4)}\frac{R^{d-4}}{\epsilon^{d-4}}+\mathcal{O}\left(\epsilon^{-(d-6)}\right)\right]+\frac{(d-2)!\Leff}{2}H_{\frac{d}{2}-1} \,.
\end{equation}
Notice that in this expression, the last term is finite and it is written in terms of the $\left(\frac{d}{2}-1\right)$-th harmonic number $H_{\frac{d}{2}-1}=\sum_{i=0}^{\frac{d}{2}-1}\frac{1}{i}$. 

Once the boundary term is computed, the Kounterterm $\SKt$ is derived easily from eq.\eqref{eq:SKt} for the even dimensional case, obtaining
\begin{equation}
    \SKt=-\frac{a_d\vol(\mathbb{S}^{d-2})\Leff^{d-1}}{4G_N}\left[\frac{1}{d-2}\frac{R^{d-2}}{\epsilon^{d-2}}-\frac{d-3}{2(d-4)}\frac{R^{d-4}}{\epsilon^{d-4}}+\mathcal{O}\left(\epsilon^{-(d-6)}\right)\right]+\frac{(d-2)!}{2}H_{\frac{d}{2}-1} \,.
\end{equation}
The structure of power-law divergences is the same as in the odd dimensional case. However, in even-dimensional CFTs, the bare EE \eqref{eq:bareSdodd} differs in a finite term and in a log term, whose coefficient is the universal part of the EE.  Following the procedure in ref.\cite{Anastasiou:2019ldc}, the log term is successfully isolated in arbitrary even $d$, reading
\begin{equation}
    S^\text{Univ}_{\text{QCG}}=(-1)^{\frac{d}{2}-1}\frac{a_d\pi^{\frac{d}{2}-1}\Leff^{d-1}}{2G_N(\frac{d}{2}-1)!}\log\frac{2R}{\epsilon} \,.
\end{equation}

\bibliographystyle{apsrev4-1}
\bibliography{entanglement}

\begin{thebibliography}{95}%
\makeatletter
\providecommand \@ifxundefined [1]{%
 \@ifx{#1\undefined}
}%
\providecommand \@ifnum [1]{%
 \ifnum #1\expandafter \@firstoftwo
 \else \expandafter \@secondoftwo
 \fi
}%
\providecommand \@ifx [1]{%
 \ifx #1\expandafter \@firstoftwo
 \else \expandafter \@secondoftwo
 \fi
}%
\providecommand \natexlab [1]{#1}%
\providecommand \enquote  [1]{``#1''}%
\providecommand \bibnamefont  [1]{#1}%
\providecommand \bibfnamefont [1]{#1}%
\providecommand \citenamefont [1]{#1}%
\providecommand \href@noop [0]{\@secondoftwo}%
\providecommand \href [0]{\begingroup \@sanitize@url \@href}%
\providecommand \@href[1]{\@@startlink{#1}\@@href}%
\providecommand \@@href[1]{\endgroup#1\@@endlink}%
\providecommand \@sanitize@url [0]{\catcode `\\12\catcode `\$12\catcode
  `\&12\catcode `\#12\catcode `\^12\catcode `\_12\catcode `\%12\relax}%
\providecommand \@@startlink[1]{}%
\providecommand \@@endlink[0]{}%
\providecommand \url  [0]{\begingroup\@sanitize@url \@url }%
\providecommand \@url [1]{\endgroup\@href {#1}{\urlprefix }}%
\providecommand \urlprefix  [0]{URL }%
\providecommand \Eprint [0]{\href }%
\providecommand \doibase [0]{http://dx.doi.org/}%
\providecommand \selectlanguage [0]{\@gobble}%
\providecommand \bibinfo  [0]{\@secondoftwo}%
\providecommand \bibfield  [0]{\@secondoftwo}%
\providecommand \translation [1]{[#1]}%
\providecommand \BibitemOpen [0]{}%
\providecommand \bibitemStop [0]{}%
\providecommand \bibitemNoStop [0]{.\EOS\space}%
\providecommand \EOS [0]{\spacefactor3000\relax}%
\providecommand \BibitemShut  [1]{\csname bibitem#1\endcsname}%
\let\auto@bib@innerbib\@empty
\bibitem [{\citenamefont {Stelle}(1977)}]{Stelle:1976gc}%
  \BibitemOpen
  \bibfield  {author} {\bibinfo {author} {\bibfnamefont {K.}~\bibnamefont
  {Stelle}},\ }\href {\doibase 10.1103/PhysRevD.16.953} {\bibfield  {journal}
  {\bibinfo  {journal} {Phys. Rev. D}\ }\textbf {\bibinfo {volume} {16}},\
  \bibinfo {pages} {953} (\bibinfo {year} {1977})}\BibitemShut {NoStop}%
\bibitem [{\citenamefont {Capper}\ and\ \citenamefont
  {Duff}(1975)}]{Capper:1975ig}%
  \BibitemOpen
  \bibfield  {author} {\bibinfo {author} {\bibfnamefont {D.}~\bibnamefont
  {Capper}}\ and\ \bibinfo {author} {\bibfnamefont {M.}~\bibnamefont {Duff}},\
  }\href {\doibase 10.1016/0375-9601(75)90030-4} {\bibfield  {journal}
  {\bibinfo  {journal} {Phys. Lett. A}\ }\textbf {\bibinfo {volume} {53}},\
  \bibinfo {pages} {361} (\bibinfo {year} {1975})}\BibitemShut {NoStop}%
\bibitem [{\citenamefont {'t~Hooft}\ and\ \citenamefont
  {Veltman}(1974)}]{tHooft:1974toh}%
  \BibitemOpen
  \bibfield  {author} {\bibinfo {author} {\bibfnamefont {G.}~\bibnamefont
  {'t~Hooft}}\ and\ \bibinfo {author} {\bibfnamefont {M.}~\bibnamefont
  {Veltman}},\ }\href@noop {} {\bibfield  {journal} {\bibinfo  {journal} {Ann.
  Inst. H. Poincare Phys. Theor. A}\ }\textbf {\bibinfo {volume} {20}},\
  \bibinfo {pages} {69} (\bibinfo {year} {1974})}\BibitemShut {NoStop}%
\bibitem [{\citenamefont {Deser}\ and\ \citenamefont {van
  Nieuwenhuizen}(1974)}]{Deser:1974cz}%
  \BibitemOpen
  \bibfield  {author} {\bibinfo {author} {\bibfnamefont {S.}~\bibnamefont
  {Deser}}\ and\ \bibinfo {author} {\bibfnamefont {P.}~\bibnamefont {van
  Nieuwenhuizen}},\ }\href {\doibase 10.1103/PhysRevD.10.401} {\bibfield
  {journal} {\bibinfo  {journal} {Phys. Rev. D}\ }\textbf {\bibinfo {volume}
  {10}},\ \bibinfo {pages} {401} (\bibinfo {year} {1974})}\BibitemShut
  {NoStop}%
\bibitem [{\citenamefont {Deser}\ \emph {et~al.}(1974)\citenamefont {Deser},
  \citenamefont {Tsao},\ and\ \citenamefont {van
  Nieuwenhuizen}}]{Deser:1974xq}%
  \BibitemOpen
  \bibfield  {author} {\bibinfo {author} {\bibfnamefont {S.}~\bibnamefont
  {Deser}}, \bibinfo {author} {\bibfnamefont {H.-S.}\ \bibnamefont {Tsao}}, \
  and\ \bibinfo {author} {\bibfnamefont {P.}~\bibnamefont {van
  Nieuwenhuizen}},\ }\href {\doibase 10.1103/PhysRevD.10.3337} {\bibfield
  {journal} {\bibinfo  {journal} {Phys. Rev. D}\ }\textbf {\bibinfo {volume}
  {10}},\ \bibinfo {pages} {3337} (\bibinfo {year} {1974})}\BibitemShut
  {NoStop}%
\bibitem [{\citenamefont {Lanczos}(1938)}]{Lanczos:1938sf}%
  \BibitemOpen
  \bibfield  {author} {\bibinfo {author} {\bibfnamefont {C.}~\bibnamefont
  {Lanczos}},\ }\href {\doibase 10.2307/1968467} {\bibfield  {journal}
  {\bibinfo  {journal} {Annals Math.}\ }\textbf {\bibinfo {volume} {39}},\
  \bibinfo {pages} {842} (\bibinfo {year} {1938})}\BibitemShut {NoStop}%
\bibitem [{\citenamefont {Lovelock}(1970)}]{lovelock1970divergence}%
  \BibitemOpen
  \bibfield  {author} {\bibinfo {author} {\bibfnamefont {D.}~\bibnamefont
  {Lovelock}},\ }\href@noop {} {\bibfield  {journal} {\bibinfo  {journal}
  {Aequationes mathematicae}\ }\textbf {\bibinfo {volume} {4}},\ \bibinfo
  {pages} {127} (\bibinfo {year} {1970})}\BibitemShut {NoStop}%
\bibitem [{\citenamefont {Lovelock}(1971)}]{lovelock1971einstein}%
  \BibitemOpen
  \bibfield  {author} {\bibinfo {author} {\bibfnamefont {D.}~\bibnamefont
  {Lovelock}},\ }\href@noop {} {\bibfield  {journal} {\bibinfo  {journal}
  {Journal of Mathematical Physics}\ }\textbf {\bibinfo {volume} {12}},\
  \bibinfo {pages} {498} (\bibinfo {year} {1971})}\BibitemShut {NoStop}%
\bibitem [{\citenamefont {Myers}\ and\ \citenamefont
  {Robinson}(2010)}]{Myers:2010ru}%
  \BibitemOpen
  \bibfield  {author} {\bibinfo {author} {\bibfnamefont {R.~C.}\ \bibnamefont
  {Myers}}\ and\ \bibinfo {author} {\bibfnamefont {B.}~\bibnamefont
  {Robinson}},\ }\href {\doibase 10.1007/JHEP08(2010)067} {\bibfield  {journal}
  {\bibinfo  {journal} {JHEP}\ }\textbf {\bibinfo {volume} {08}},\ \bibinfo
  {pages} {067} (\bibinfo {year} {2010})},\ \Eprint
  {http://arxiv.org/abs/1003.5357} {arXiv:1003.5357 [gr-qc]} \BibitemShut
  {NoStop}%
\bibitem [{\citenamefont {Oliva}\ and\ \citenamefont
  {Ray}(2010)}]{Oliva:2010eb}%
  \BibitemOpen
  \bibfield  {author} {\bibinfo {author} {\bibfnamefont {J.}~\bibnamefont
  {Oliva}}\ and\ \bibinfo {author} {\bibfnamefont {S.}~\bibnamefont {Ray}},\
  }\href {\doibase 10.1088/0264-9381/27/22/225002} {\bibfield  {journal}
  {\bibinfo  {journal} {Class. Quant. Grav.}\ }\textbf {\bibinfo {volume}
  {27}},\ \bibinfo {pages} {225002} (\bibinfo {year} {2010})},\ \Eprint
  {http://arxiv.org/abs/1003.4773} {arXiv:1003.4773 [gr-qc]} \BibitemShut
  {NoStop}%
\bibitem [{\citenamefont {Bueno}\ and\ \citenamefont
  {Cano}(2016)}]{Bueno:2016xff}%
  \BibitemOpen
  \bibfield  {author} {\bibinfo {author} {\bibfnamefont {P.}~\bibnamefont
  {Bueno}}\ and\ \bibinfo {author} {\bibfnamefont {P.~A.}\ \bibnamefont
  {Cano}},\ }\href {\doibase 10.1103/PhysRevD.94.104005} {\bibfield  {journal}
  {\bibinfo  {journal} {Phys. Rev. D}\ }\textbf {\bibinfo {volume} {94}},\
  \bibinfo {pages} {104005} (\bibinfo {year} {2016})},\ \Eprint
  {http://arxiv.org/abs/1607.06463} {arXiv:1607.06463 [hep-th]} \BibitemShut
  {NoStop}%
\bibitem [{\citenamefont {Hennigar}\ \emph {et~al.}(2017)\citenamefont
  {Hennigar}, \citenamefont {Kubiz\v{n}\'ak},\ and\ \citenamefont
  {Mann}}]{Hennigar:2017ego}%
  \BibitemOpen
  \bibfield  {author} {\bibinfo {author} {\bibfnamefont {R.~A.}\ \bibnamefont
  {Hennigar}}, \bibinfo {author} {\bibfnamefont {D.}~\bibnamefont
  {Kubiz\v{n}\'ak}}, \ and\ \bibinfo {author} {\bibfnamefont {R.~B.}\
  \bibnamefont {Mann}},\ }\href {\doibase 10.1103/PhysRevD.95.104042}
  {\bibfield  {journal} {\bibinfo  {journal} {Phys. Rev. D}\ }\textbf {\bibinfo
  {volume} {95}},\ \bibinfo {pages} {104042} (\bibinfo {year} {2017})},\
  \Eprint {http://arxiv.org/abs/1703.01631} {arXiv:1703.01631 [hep-th]}
  \BibitemShut {NoStop}%
\bibitem [{\citenamefont {Bueno}\ \emph
  {et~al.}(2019{\natexlab{a}})\citenamefont {Bueno}, \citenamefont {Cano},
  \citenamefont {Moreno},\ and\ \citenamefont {Murcia}}]{Bueno:2019ltp}%
  \BibitemOpen
  \bibfield  {author} {\bibinfo {author} {\bibfnamefont {P.}~\bibnamefont
  {Bueno}}, \bibinfo {author} {\bibfnamefont {P.~A.}\ \bibnamefont {Cano}},
  \bibinfo {author} {\bibfnamefont {J.}~\bibnamefont {Moreno}}, \ and\ \bibinfo
  {author} {\bibfnamefont {A.}~\bibnamefont {Murcia}},\ }\href {\doibase
  10.1007/JHEP11(2019)062} {\bibfield  {journal} {\bibinfo  {journal} {JHEP}\
  }\textbf {\bibinfo {volume} {11}},\ \bibinfo {pages} {062} (\bibinfo {year}
  {2019}{\natexlab{a}})},\ \Eprint {http://arxiv.org/abs/1906.00987}
  {arXiv:1906.00987 [hep-th]} \BibitemShut {NoStop}%
\bibitem [{\citenamefont {Bueno}\ \emph
  {et~al.}(2020{\natexlab{a}})\citenamefont {Bueno}, \citenamefont {Cano},\
  and\ \citenamefont {Hennigar}}]{Bueno:2019ycr}%
  \BibitemOpen
  \bibfield  {author} {\bibinfo {author} {\bibfnamefont {P.}~\bibnamefont
  {Bueno}}, \bibinfo {author} {\bibfnamefont {P.~A.}\ \bibnamefont {Cano}}, \
  and\ \bibinfo {author} {\bibfnamefont {R.~A.}\ \bibnamefont {Hennigar}},\
  }\href {\doibase 10.1088/1361-6382/ab5410} {\bibfield  {journal} {\bibinfo
  {journal} {Class. Quant. Grav.}\ }\textbf {\bibinfo {volume} {37}},\ \bibinfo
  {pages} {015002} (\bibinfo {year} {2020}{\natexlab{a}})},\ \Eprint
  {http://arxiv.org/abs/1909.07983} {arXiv:1909.07983 [hep-th]} \BibitemShut
  {NoStop}%
\bibitem [{\citenamefont {Lu}\ and\ \citenamefont {Pope}(2011)}]{Lu:2011zk}%
  \BibitemOpen
  \bibfield  {author} {\bibinfo {author} {\bibfnamefont {H.}~\bibnamefont
  {Lu}}\ and\ \bibinfo {author} {\bibfnamefont {C.~N.}\ \bibnamefont {Pope}},\
  }\href {\doibase 10.1103/PhysRevLett.106.181302} {\bibfield  {journal}
  {\bibinfo  {journal} {Phys. Rev. Lett.}\ }\textbf {\bibinfo {volume} {106}},\
  \bibinfo {pages} {181302} (\bibinfo {year} {2011})},\ \Eprint
  {http://arxiv.org/abs/1101.1971} {arXiv:1101.1971 [hep-th]} \BibitemShut
  {NoStop}%
\bibitem [{\citenamefont {Anastasiou}\ \emph
  {et~al.}(2019{\natexlab{a}})\citenamefont {Anastasiou}, \citenamefont
  {Olea},\ and\ \citenamefont {Rivera-Betancour}}]{Anastasiou:2017rjf}%
  \BibitemOpen
  \bibfield  {author} {\bibinfo {author} {\bibfnamefont {G.}~\bibnamefont
  {Anastasiou}}, \bibinfo {author} {\bibfnamefont {R.}~\bibnamefont {Olea}}, \
  and\ \bibinfo {author} {\bibfnamefont {D.}~\bibnamefont {Rivera-Betancour}},\
  }\href {\doibase 10.1016/j.physletb.2018.11.021} {\bibfield  {journal}
  {\bibinfo  {journal} {Phys. Lett. B}\ }\textbf {\bibinfo {volume} {788}},\
  \bibinfo {pages} {302} (\bibinfo {year} {2019}{\natexlab{a}})},\ \Eprint
  {http://arxiv.org/abs/1707.00341} {arXiv:1707.00341 [hep-th]} \BibitemShut
  {NoStop}%
\bibitem [{\citenamefont {Maldacena}(1999)}]{Maldacena:1997re}%
  \BibitemOpen
  \bibfield  {author} {\bibinfo {author} {\bibfnamefont {J.~M.}\ \bibnamefont
  {Maldacena}},\ }\href {\doibase 10.1023/A:1026654312961} {\bibfield
  {journal} {\bibinfo  {journal} {Int. J. Theor. Phys.}\ }\textbf {\bibinfo
  {volume} {38}},\ \bibinfo {pages} {1113} (\bibinfo {year} {1999})},\ \Eprint
  {http://arxiv.org/abs/hep-th/9711200} {arXiv:hep-th/9711200} \BibitemShut
  {NoStop}%
\bibitem [{\citenamefont {Gubser}\ \emph {et~al.}(1998)\citenamefont {Gubser},
  \citenamefont {Klebanov},\ and\ \citenamefont {Polyakov}}]{Gubser:1998bc}%
  \BibitemOpen
  \bibfield  {author} {\bibinfo {author} {\bibfnamefont {S.}~\bibnamefont
  {Gubser}}, \bibinfo {author} {\bibfnamefont {I.~R.}\ \bibnamefont
  {Klebanov}}, \ and\ \bibinfo {author} {\bibfnamefont {A.~M.}\ \bibnamefont
  {Polyakov}},\ }\href {\doibase 10.1016/S0370-2693(98)00377-3} {\bibfield
  {journal} {\bibinfo  {journal} {Phys. Lett. B}\ }\textbf {\bibinfo {volume}
  {428}},\ \bibinfo {pages} {105} (\bibinfo {year} {1998})},\ \Eprint
  {http://arxiv.org/abs/hep-th/9802109} {arXiv:hep-th/9802109} \BibitemShut
  {NoStop}%
\bibitem [{\citenamefont {Witten}(1998)}]{Witten:1998qj}%
  \BibitemOpen
  \bibfield  {author} {\bibinfo {author} {\bibfnamefont {E.}~\bibnamefont
  {Witten}},\ }\href {\doibase 10.4310/ATMP.1998.v2.n2.a2} {\bibfield
  {journal} {\bibinfo  {journal} {Adv. Theor. Math. Phys.}\ }\textbf {\bibinfo
  {volume} {2}},\ \bibinfo {pages} {253} (\bibinfo {year} {1998})},\ \Eprint
  {http://arxiv.org/abs/hep-th/9802150} {arXiv:hep-th/9802150} \BibitemShut
  {NoStop}%
\bibitem [{\citenamefont {Nojiri}\ and\ \citenamefont
  {Odintsov}(2000)}]{Nojiri:1999mh}%
  \BibitemOpen
  \bibfield  {author} {\bibinfo {author} {\bibfnamefont {S.}~\bibnamefont
  {Nojiri}}\ and\ \bibinfo {author} {\bibfnamefont {S.~D.}\ \bibnamefont
  {Odintsov}},\ }\href {\doibase 10.1142/S0217751X00000197} {\bibfield
  {journal} {\bibinfo  {journal} {Int. J. Mod. Phys. A}\ }\textbf {\bibinfo
  {volume} {15}},\ \bibinfo {pages} {413} (\bibinfo {year} {2000})},\ \Eprint
  {http://arxiv.org/abs/hep-th/9903033} {arXiv:hep-th/9903033} \BibitemShut
  {NoStop}%
\bibitem [{\citenamefont {Ghodsi}\ and\ \citenamefont
  {Siahvoshan}(2019)}]{Ghodsi:2019xrx}%
  \BibitemOpen
  \bibfield  {author} {\bibinfo {author} {\bibfnamefont {A.}~\bibnamefont
  {Ghodsi}}\ and\ \bibinfo {author} {\bibfnamefont {M.}~\bibnamefont
  {Siahvoshan}},\ }\href {\doibase 10.1140/epjc/s10052-019-7345-8} {\bibfield
  {journal} {\bibinfo  {journal} {Eur. Phys. J. C}\ }\textbf {\bibinfo {volume}
  {79}},\ \bibinfo {pages} {820} (\bibinfo {year} {2019})},\ \Eprint
  {http://arxiv.org/abs/1907.03497} {arXiv:1907.03497 [hep-th]} \BibitemShut
  {NoStop}%
\bibitem [{\citenamefont {Policastro}\ \emph {et~al.}(2001)\citenamefont
  {Policastro}, \citenamefont {Son},\ and\ \citenamefont
  {Starinets}}]{Policastro:2001yc}%
  \BibitemOpen
  \bibfield  {author} {\bibinfo {author} {\bibfnamefont {G.}~\bibnamefont
  {Policastro}}, \bibinfo {author} {\bibfnamefont {D.~T.}\ \bibnamefont {Son}},
  \ and\ \bibinfo {author} {\bibfnamefont {A.~O.}\ \bibnamefont {Starinets}},\
  }\href {\doibase 10.1103/PhysRevLett.87.081601} {\bibfield  {journal}
  {\bibinfo  {journal} {Phys. Rev. Lett.}\ }\textbf {\bibinfo {volume} {87}},\
  \bibinfo {pages} {081601} (\bibinfo {year} {2001})},\ \Eprint
  {http://arxiv.org/abs/hep-th/0104066} {arXiv:hep-th/0104066} \BibitemShut
  {NoStop}%
\bibitem [{\citenamefont {Buchel}\ and\ \citenamefont
  {Liu}(2004)}]{Buchel:2003tz}%
  \BibitemOpen
  \bibfield  {author} {\bibinfo {author} {\bibfnamefont {A.}~\bibnamefont
  {Buchel}}\ and\ \bibinfo {author} {\bibfnamefont {J.~T.}\ \bibnamefont
  {Liu}},\ }\href {\doibase 10.1103/PhysRevLett.93.090602} {\bibfield
  {journal} {\bibinfo  {journal} {Phys. Rev. Lett.}\ }\textbf {\bibinfo
  {volume} {93}},\ \bibinfo {pages} {090602} (\bibinfo {year} {2004})},\
  \Eprint {http://arxiv.org/abs/hep-th/0311175} {arXiv:hep-th/0311175}
  \BibitemShut {NoStop}%
\bibitem [{\citenamefont {Kovtun}\ \emph {et~al.}(2005)\citenamefont {Kovtun},
  \citenamefont {Son},\ and\ \citenamefont {Starinets}}]{Kovtun:2004de}%
  \BibitemOpen
  \bibfield  {author} {\bibinfo {author} {\bibfnamefont {P.}~\bibnamefont
  {Kovtun}}, \bibinfo {author} {\bibfnamefont {D.~T.}\ \bibnamefont {Son}}, \
  and\ \bibinfo {author} {\bibfnamefont {A.~O.}\ \bibnamefont {Starinets}},\
  }\href {\doibase 10.1103/PhysRevLett.94.111601} {\bibfield  {journal}
  {\bibinfo  {journal} {Phys. Rev. Lett.}\ }\textbf {\bibinfo {volume} {94}},\
  \bibinfo {pages} {111601} (\bibinfo {year} {2005})},\ \Eprint
  {http://arxiv.org/abs/hep-th/0405231} {arXiv:hep-th/0405231} \BibitemShut
  {NoStop}%
\bibitem [{\citenamefont {Brigante}\ \emph
  {et~al.}(2008{\natexlab{a}})\citenamefont {Brigante}, \citenamefont {Liu},
  \citenamefont {Myers}, \citenamefont {Shenker},\ and\ \citenamefont
  {Yaida}}]{Brigante:2007nu}%
  \BibitemOpen
  \bibfield  {author} {\bibinfo {author} {\bibfnamefont {M.}~\bibnamefont
  {Brigante}}, \bibinfo {author} {\bibfnamefont {H.}~\bibnamefont {Liu}},
  \bibinfo {author} {\bibfnamefont {R.~C.}\ \bibnamefont {Myers}}, \bibinfo
  {author} {\bibfnamefont {S.}~\bibnamefont {Shenker}}, \ and\ \bibinfo
  {author} {\bibfnamefont {S.}~\bibnamefont {Yaida}},\ }\href {\doibase
  10.1103/PhysRevD.77.126006} {\bibfield  {journal} {\bibinfo  {journal} {Phys.
  Rev. D}\ }\textbf {\bibinfo {volume} {77}},\ \bibinfo {pages} {126006}
  (\bibinfo {year} {2008}{\natexlab{a}})},\ \Eprint
  {http://arxiv.org/abs/0712.0805} {arXiv:0712.0805 [hep-th]} \BibitemShut
  {NoStop}%
\bibitem [{\citenamefont {Camanho}\ and\ \citenamefont
  {Edelstein}(2010)}]{Camanho:2009vw}%
  \BibitemOpen
  \bibfield  {author} {\bibinfo {author} {\bibfnamefont {X.~O.}\ \bibnamefont
  {Camanho}}\ and\ \bibinfo {author} {\bibfnamefont {J.~D.}\ \bibnamefont
  {Edelstein}},\ }\href {\doibase 10.1007/JHEP04(2010)007} {\bibfield
  {journal} {\bibinfo  {journal} {JHEP}\ }\textbf {\bibinfo {volume} {04}},\
  \bibinfo {pages} {007} (\bibinfo {year} {2010})},\ \Eprint
  {http://arxiv.org/abs/0911.3160} {arXiv:0911.3160 [hep-th]} \BibitemShut
  {NoStop}%
\bibitem [{\citenamefont {Brigante}\ \emph
  {et~al.}(2008{\natexlab{b}})\citenamefont {Brigante}, \citenamefont {Liu},
  \citenamefont {Myers}, \citenamefont {Shenker},\ and\ \citenamefont
  {Yaida}}]{Brigante:2008gz}%
  \BibitemOpen
  \bibfield  {author} {\bibinfo {author} {\bibfnamefont {M.}~\bibnamefont
  {Brigante}}, \bibinfo {author} {\bibfnamefont {H.}~\bibnamefont {Liu}},
  \bibinfo {author} {\bibfnamefont {R.~C.}\ \bibnamefont {Myers}}, \bibinfo
  {author} {\bibfnamefont {S.}~\bibnamefont {Shenker}}, \ and\ \bibinfo
  {author} {\bibfnamefont {S.}~\bibnamefont {Yaida}},\ }\href {\doibase
  10.1103/PhysRevLett.100.191601} {\bibfield  {journal} {\bibinfo  {journal}
  {Phys. Rev. Lett.}\ }\textbf {\bibinfo {volume} {100}},\ \bibinfo {pages}
  {191601} (\bibinfo {year} {2008}{\natexlab{b}})},\ \Eprint
  {http://arxiv.org/abs/0802.3318} {arXiv:0802.3318 [hep-th]} \BibitemShut
  {NoStop}%
\bibitem [{\citenamefont {de~Boer}\ \emph {et~al.}(2010)\citenamefont
  {de~Boer}, \citenamefont {Kulaxizi},\ and\ \citenamefont
  {Parnachev}}]{deBoer:2009pn}%
  \BibitemOpen
  \bibfield  {author} {\bibinfo {author} {\bibfnamefont {J.}~\bibnamefont
  {de~Boer}}, \bibinfo {author} {\bibfnamefont {M.}~\bibnamefont {Kulaxizi}}, \
  and\ \bibinfo {author} {\bibfnamefont {A.}~\bibnamefont {Parnachev}},\ }\href
  {\doibase 10.1007/JHEP03(2010)087} {\bibfield  {journal} {\bibinfo  {journal}
  {JHEP}\ }\textbf {\bibinfo {volume} {03}},\ \bibinfo {pages} {087} (\bibinfo
  {year} {2010})},\ \Eprint {http://arxiv.org/abs/0910.5347} {arXiv:0910.5347
  [hep-th]} \BibitemShut {NoStop}%
\bibitem [{\citenamefont {Camanho}\ \emph {et~al.}(2011)\citenamefont
  {Camanho}, \citenamefont {Edelstein},\ and\ \citenamefont
  {Paulos}}]{Camanho:2010ru}%
  \BibitemOpen
  \bibfield  {author} {\bibinfo {author} {\bibfnamefont {X.~O.}\ \bibnamefont
  {Camanho}}, \bibinfo {author} {\bibfnamefont {J.~D.}\ \bibnamefont
  {Edelstein}}, \ and\ \bibinfo {author} {\bibfnamefont {M.~F.}\ \bibnamefont
  {Paulos}},\ }\href {\doibase 10.1007/JHEP05(2011)127} {\bibfield  {journal}
  {\bibinfo  {journal} {JHEP}\ }\textbf {\bibinfo {volume} {05}},\ \bibinfo
  {pages} {127} (\bibinfo {year} {2011})},\ \Eprint
  {http://arxiv.org/abs/1010.1682} {arXiv:1010.1682 [hep-th]} \BibitemShut
  {NoStop}%
\bibitem [{\citenamefont {Myers}\ and\ \citenamefont
  {Sinha}(2011)}]{Myers:2010tj}%
  \BibitemOpen
  \bibfield  {author} {\bibinfo {author} {\bibfnamefont {R.~C.}\ \bibnamefont
  {Myers}}\ and\ \bibinfo {author} {\bibfnamefont {A.}~\bibnamefont {Sinha}},\
  }\href {\doibase 10.1007/JHEP01(2011)125} {\bibfield  {journal} {\bibinfo
  {journal} {JHEP}\ }\textbf {\bibinfo {volume} {01}},\ \bibinfo {pages} {125}
  (\bibinfo {year} {2011})},\ \Eprint {http://arxiv.org/abs/1011.5819}
  {arXiv:1011.5819 [hep-th]} \BibitemShut {NoStop}%
\bibitem [{\citenamefont {Myers}\ and\ \citenamefont
  {Sinha}(2010)}]{Myers:2010xs}%
  \BibitemOpen
  \bibfield  {author} {\bibinfo {author} {\bibfnamefont {R.~C.}\ \bibnamefont
  {Myers}}\ and\ \bibinfo {author} {\bibfnamefont {A.}~\bibnamefont {Sinha}},\
  }\href {\doibase 10.1103/PhysRevD.82.046006} {\bibfield  {journal} {\bibinfo
  {journal} {Phys. Rev.}\ }\textbf {\bibinfo {volume} {D82}},\ \bibinfo {pages}
  {046006} (\bibinfo {year} {2010})},\ \Eprint {http://arxiv.org/abs/1006.1263}
  {arXiv:1006.1263 [hep-th]} \BibitemShut {NoStop}%
\bibitem [{\citenamefont {Mezei}(2015)}]{Mezei:2014zla}%
  \BibitemOpen
  \bibfield  {author} {\bibinfo {author} {\bibfnamefont {M.}~\bibnamefont
  {Mezei}},\ }\href {\doibase 10.1103/PhysRevD.91.045038} {\bibfield  {journal}
  {\bibinfo  {journal} {Phys. Rev.}\ }\textbf {\bibinfo {volume} {D91}},\
  \bibinfo {pages} {045038} (\bibinfo {year} {2015})},\ \Eprint
  {http://arxiv.org/abs/1411.7011} {arXiv:1411.7011 [hep-th]} \BibitemShut
  {NoStop}%
\bibitem [{\citenamefont {Bueno}\ \emph
  {et~al.}(2019{\natexlab{b}})\citenamefont {Bueno}, \citenamefont {Cano},
  \citenamefont {Hennigar},\ and\ \citenamefont {Mann}}]{Bueno:2018yzo}%
  \BibitemOpen
  \bibfield  {author} {\bibinfo {author} {\bibfnamefont {P.}~\bibnamefont
  {Bueno}}, \bibinfo {author} {\bibfnamefont {P.~A.}\ \bibnamefont {Cano}},
  \bibinfo {author} {\bibfnamefont {R.~A.}\ \bibnamefont {Hennigar}}, \ and\
  \bibinfo {author} {\bibfnamefont {R.~B.}\ \bibnamefont {Mann}},\ }\href
  {\doibase 10.1103/PhysRevLett.122.071602} {\bibfield  {journal} {\bibinfo
  {journal} {Phys. Rev. Lett.}\ }\textbf {\bibinfo {volume} {122}},\ \bibinfo
  {pages} {071602} (\bibinfo {year} {2019}{\natexlab{b}})},\ \Eprint
  {http://arxiv.org/abs/1808.02052} {arXiv:1808.02052 [hep-th]} \BibitemShut
  {NoStop}%
\bibitem [{\citenamefont {Bueno}\ \emph
  {et~al.}(2020{\natexlab{b}})\citenamefont {Bueno}, \citenamefont {Cano},
  \citenamefont {Hennigar}, \citenamefont {Penas},\ and\ \citenamefont
  {Ruip\'erez}}]{Bueno:2020odt}%
  \BibitemOpen
  \bibfield  {author} {\bibinfo {author} {\bibfnamefont {P.}~\bibnamefont
  {Bueno}}, \bibinfo {author} {\bibfnamefont {P.~A.}\ \bibnamefont {Cano}},
  \bibinfo {author} {\bibfnamefont {R.~A.}\ \bibnamefont {Hennigar}}, \bibinfo
  {author} {\bibfnamefont {V.~A.}\ \bibnamefont {Penas}}, \ and\ \bibinfo
  {author} {\bibfnamefont {A.}~\bibnamefont {Ruip\'erez}},\ }\href {\doibase
  10.1007/JHEP04(2020)123} {\bibfield  {journal} {\bibinfo  {journal} {JHEP}\
  }\textbf {\bibinfo {volume} {04}},\ \bibinfo {pages} {123} (\bibinfo {year}
  {2020}{\natexlab{b}})},\ \Eprint {http://arxiv.org/abs/2001.10020}
  {arXiv:2001.10020 [hep-th]} \BibitemShut {NoStop}%
\bibitem [{\citenamefont {Hung}\ \emph {et~al.}(2011)\citenamefont {Hung},
  \citenamefont {Myers},\ and\ \citenamefont {Smolkin}}]{Hung:2011xb}%
  \BibitemOpen
  \bibfield  {author} {\bibinfo {author} {\bibfnamefont {L.-Y.}\ \bibnamefont
  {Hung}}, \bibinfo {author} {\bibfnamefont {R.~C.}\ \bibnamefont {Myers}}, \
  and\ \bibinfo {author} {\bibfnamefont {M.}~\bibnamefont {Smolkin}},\ }\href
  {\doibase 10.1007/JHEP04(2011)025} {\bibfield  {journal} {\bibinfo  {journal}
  {JHEP}\ }\textbf {\bibinfo {volume} {04}},\ \bibinfo {pages} {025} (\bibinfo
  {year} {2011})},\ \Eprint {http://arxiv.org/abs/1101.5813} {arXiv:1101.5813
  [hep-th]} \BibitemShut {NoStop}%
\bibitem [{\citenamefont {Grover}\ \emph {et~al.}(2011)\citenamefont {Grover},
  \citenamefont {Turner},\ and\ \citenamefont {Vishwanath}}]{Grover:2011fa}%
  \BibitemOpen
  \bibfield  {author} {\bibinfo {author} {\bibfnamefont {T.}~\bibnamefont
  {Grover}}, \bibinfo {author} {\bibfnamefont {A.~M.}\ \bibnamefont {Turner}},
  \ and\ \bibinfo {author} {\bibfnamefont {A.}~\bibnamefont {Vishwanath}},\
  }\href {\doibase 10.1103/PhysRevB.84.195120} {\bibfield  {journal} {\bibinfo
  {journal} {Phys. Rev.}\ }\textbf {\bibinfo {volume} {B84}},\ \bibinfo {pages}
  {195120} (\bibinfo {year} {2011})},\ \Eprint {http://arxiv.org/abs/1108.4038}
  {arXiv:1108.4038 [cond-mat.str-el]} \BibitemShut {NoStop}%
\bibitem [{\citenamefont {Myers}\ and\ \citenamefont
  {Singh}(2012)}]{Myers:2012ed}%
  \BibitemOpen
  \bibfield  {author} {\bibinfo {author} {\bibfnamefont {R.~C.}\ \bibnamefont
  {Myers}}\ and\ \bibinfo {author} {\bibfnamefont {A.}~\bibnamefont {Singh}},\
  }\href {\doibase 10.1007/JHEP04(2012)122} {\bibfield  {journal} {\bibinfo
  {journal} {JHEP}\ }\textbf {\bibinfo {volume} {04}},\ \bibinfo {pages} {122}
  (\bibinfo {year} {2012})},\ \Eprint {http://arxiv.org/abs/1202.2068}
  {arXiv:1202.2068 [hep-th]} \BibitemShut {NoStop}%
\bibitem [{\citenamefont {Liu}\ and\ \citenamefont
  {Mezei}(2013)}]{Liu:2012eea}%
  \BibitemOpen
  \bibfield  {author} {\bibinfo {author} {\bibfnamefont {H.}~\bibnamefont
  {Liu}}\ and\ \bibinfo {author} {\bibfnamefont {M.}~\bibnamefont {Mezei}},\
  }\href {\doibase 10.1007/JHEP04(2013)162} {\bibfield  {journal} {\bibinfo
  {journal} {JHEP}\ }\textbf {\bibinfo {volume} {04}},\ \bibinfo {pages} {162}
  (\bibinfo {year} {2013})},\ \Eprint {http://arxiv.org/abs/1202.2070}
  {arXiv:1202.2070 [hep-th]} \BibitemShut {NoStop}%
\bibitem [{\citenamefont {Jafferis}\ \emph {et~al.}(2011)\citenamefont
  {Jafferis}, \citenamefont {Klebanov}, \citenamefont {Pufu},\ and\
  \citenamefont {Safdi}}]{Jafferis:2011zi}%
  \BibitemOpen
  \bibfield  {author} {\bibinfo {author} {\bibfnamefont {D.~L.}\ \bibnamefont
  {Jafferis}}, \bibinfo {author} {\bibfnamefont {I.~R.}\ \bibnamefont
  {Klebanov}}, \bibinfo {author} {\bibfnamefont {S.~S.}\ \bibnamefont {Pufu}},
  \ and\ \bibinfo {author} {\bibfnamefont {B.~R.}\ \bibnamefont {Safdi}},\
  }\href {\doibase 10.1007/JHEP06(2011)102} {\bibfield  {journal} {\bibinfo
  {journal} {JHEP}\ }\textbf {\bibinfo {volume} {06}},\ \bibinfo {pages} {102}
  (\bibinfo {year} {2011})},\ \Eprint {http://arxiv.org/abs/1103.1181}
  {arXiv:1103.1181 [hep-th]} \BibitemShut {NoStop}%
\bibitem [{\citenamefont {Klebanov}\ \emph {et~al.}(2011)\citenamefont
  {Klebanov}, \citenamefont {Pufu},\ and\ \citenamefont
  {Safdi}}]{Klebanov:2011gs}%
  \BibitemOpen
  \bibfield  {author} {\bibinfo {author} {\bibfnamefont {I.~R.}\ \bibnamefont
  {Klebanov}}, \bibinfo {author} {\bibfnamefont {S.~S.}\ \bibnamefont {Pufu}},
  \ and\ \bibinfo {author} {\bibfnamefont {B.~R.}\ \bibnamefont {Safdi}},\
  }\href {\doibase 10.1007/JHEP10(2011)038} {\bibfield  {journal} {\bibinfo
  {journal} {JHEP}\ }\textbf {\bibinfo {volume} {10}},\ \bibinfo {pages} {038}
  (\bibinfo {year} {2011})},\ \Eprint {http://arxiv.org/abs/1105.4598}
  {arXiv:1105.4598 [hep-th]} \BibitemShut {NoStop}%
\bibitem [{\citenamefont {Casini}\ \emph {et~al.}(2011)\citenamefont {Casini},
  \citenamefont {Huerta},\ and\ \citenamefont {Myers}}]{Casini:2011kv}%
  \BibitemOpen
  \bibfield  {author} {\bibinfo {author} {\bibfnamefont {H.}~\bibnamefont
  {Casini}}, \bibinfo {author} {\bibfnamefont {M.}~\bibnamefont {Huerta}}, \
  and\ \bibinfo {author} {\bibfnamefont {R.~C.}\ \bibnamefont {Myers}},\ }\href
  {\doibase 10.1007/JHEP05(2011)036} {\bibfield  {journal} {\bibinfo  {journal}
  {JHEP}\ }\textbf {\bibinfo {volume} {05}},\ \bibinfo {pages} {036} (\bibinfo
  {year} {2011})},\ \Eprint {http://arxiv.org/abs/1102.0440} {arXiv:1102.0440
  [hep-th]} \BibitemShut {NoStop}%
\bibitem [{\citenamefont {Zamolodchikov}(1986)}]{Zamolodchikov:1986gt}%
  \BibitemOpen
  \bibfield  {author} {\bibinfo {author} {\bibfnamefont {A.~B.}\ \bibnamefont
  {Zamolodchikov}},\ }\href@noop {} {\bibfield  {journal} {\bibinfo  {journal}
  {JETP Lett.}\ }\textbf {\bibinfo {volume} {43}},\ \bibinfo {pages} {730}
  (\bibinfo {year} {1986})}\BibitemShut {NoStop}%
\bibitem [{\citenamefont {Cardy}(1988)}]{Cardy:1988cwa}%
  \BibitemOpen
  \bibfield  {author} {\bibinfo {author} {\bibfnamefont {J.~L.}\ \bibnamefont
  {Cardy}},\ }\href {\doibase 10.1016/0370-2693(88)90054-8} {\bibfield
  {journal} {\bibinfo  {journal} {Phys. Lett. B}\ }\textbf {\bibinfo {volume}
  {215}},\ \bibinfo {pages} {749} (\bibinfo {year} {1988})}\BibitemShut
  {NoStop}%
\bibitem [{\citenamefont {Komargodski}\ and\ \citenamefont
  {Schwimmer}(2011)}]{Komargodski:2011vj}%
  \BibitemOpen
  \bibfield  {author} {\bibinfo {author} {\bibfnamefont {Z.}~\bibnamefont
  {Komargodski}}\ and\ \bibinfo {author} {\bibfnamefont {A.}~\bibnamefont
  {Schwimmer}},\ }\href {\doibase 10.1007/JHEP12(2011)099} {\bibfield
  {journal} {\bibinfo  {journal} {JHEP}\ }\textbf {\bibinfo {volume} {12}},\
  \bibinfo {pages} {099} (\bibinfo {year} {2011})},\ \Eprint
  {http://arxiv.org/abs/1107.3987} {arXiv:1107.3987 [hep-th]} \BibitemShut
  {NoStop}%
\bibitem [{\citenamefont {Calabrese}\ and\ \citenamefont
  {Cardy}(2004)}]{Calabrese:2004eu}%
  \BibitemOpen
  \bibfield  {author} {\bibinfo {author} {\bibfnamefont {P.}~\bibnamefont
  {Calabrese}}\ and\ \bibinfo {author} {\bibfnamefont {J.~L.}\ \bibnamefont
  {Cardy}},\ }\href {\doibase 10.1088/1742-5468/2004/06/P06002} {\bibfield
  {journal} {\bibinfo  {journal} {J. Stat. Mech.}\ }\textbf {\bibinfo {volume}
  {0406}},\ \bibinfo {pages} {P06002} (\bibinfo {year} {2004})},\ \Eprint
  {http://arxiv.org/abs/hep-th/0405152} {arXiv:hep-th/0405152 [hep-th]}
  \BibitemShut {NoStop}%
\bibitem [{\citenamefont {Ryu}\ and\ \citenamefont
  {Takayanagi}(2006{\natexlab{a}})}]{Ryu:2006bv}%
  \BibitemOpen
  \bibfield  {author} {\bibinfo {author} {\bibfnamefont {S.}~\bibnamefont
  {Ryu}}\ and\ \bibinfo {author} {\bibfnamefont {T.}~\bibnamefont
  {Takayanagi}},\ }\href {\doibase 10.1103/PhysRevLett.96.181602} {\bibfield
  {journal} {\bibinfo  {journal} {Phys. Rev. Lett.}\ }\textbf {\bibinfo
  {volume} {96}},\ \bibinfo {pages} {181602} (\bibinfo {year}
  {2006}{\natexlab{a}})},\ \Eprint {http://arxiv.org/abs/hep-th/0603001}
  {arXiv:hep-th/0603001 [hep-th]} \BibitemShut {NoStop}%
\bibitem [{\citenamefont {Ryu}\ and\ \citenamefont
  {Takayanagi}(2006{\natexlab{b}})}]{Ryu:2006ef}%
  \BibitemOpen
  \bibfield  {author} {\bibinfo {author} {\bibfnamefont {S.}~\bibnamefont
  {Ryu}}\ and\ \bibinfo {author} {\bibfnamefont {T.}~\bibnamefont
  {Takayanagi}},\ }\href {\doibase 10.1088/1126-6708/2006/08/045} {\bibfield
  {journal} {\bibinfo  {journal} {JHEP}\ }\textbf {\bibinfo {volume} {08}},\
  \bibinfo {pages} {045} (\bibinfo {year} {2006}{\natexlab{b}})},\ \Eprint
  {http://arxiv.org/abs/hep-th/0605073} {arXiv:hep-th/0605073 [hep-th]}
  \BibitemShut {NoStop}%
\bibitem [{\citenamefont {Lewkowycz}\ and\ \citenamefont
  {Maldacena}(2013)}]{Lewkowycz:2013nqa}%
  \BibitemOpen
  \bibfield  {author} {\bibinfo {author} {\bibfnamefont {A.}~\bibnamefont
  {Lewkowycz}}\ and\ \bibinfo {author} {\bibfnamefont {J.}~\bibnamefont
  {Maldacena}},\ }\href {\doibase 10.1007/JHEP08(2013)090} {\bibfield
  {journal} {\bibinfo  {journal} {JHEP}\ }\textbf {\bibinfo {volume} {08}},\
  \bibinfo {pages} {090} (\bibinfo {year} {2013})},\ \Eprint
  {http://arxiv.org/abs/1304.4926} {arXiv:1304.4926 [hep-th]} \BibitemShut
  {NoStop}%
\bibitem [{\citenamefont {Nishioka}(2018)}]{Nishioka:2018khk}%
  \BibitemOpen
  \bibfield  {author} {\bibinfo {author} {\bibfnamefont {T.}~\bibnamefont
  {Nishioka}},\ }\href {\doibase 10.1103/RevModPhys.90.035007} {\bibfield
  {journal} {\bibinfo  {journal} {Rev. Mod. Phys.}\ }\textbf {\bibinfo {volume}
  {90}},\ \bibinfo {pages} {035007} (\bibinfo {year} {2018})},\ \Eprint
  {http://arxiv.org/abs/1801.10352} {arXiv:1801.10352 [hep-th]} \BibitemShut
  {NoStop}%
\bibitem [{\citenamefont {Fursaev}\ \emph {et~al.}(2013)\citenamefont
  {Fursaev}, \citenamefont {Patrushev},\ and\ \citenamefont
  {Solodukhin}}]{Fursaev:2013fta}%
  \BibitemOpen
  \bibfield  {author} {\bibinfo {author} {\bibfnamefont {D.~V.}\ \bibnamefont
  {Fursaev}}, \bibinfo {author} {\bibfnamefont {A.}~\bibnamefont {Patrushev}},
  \ and\ \bibinfo {author} {\bibfnamefont {S.~N.}\ \bibnamefont {Solodukhin}},\
  }\href {\doibase 10.1103/PhysRevD.88.044054} {\bibfield  {journal} {\bibinfo
  {journal} {Phys. Rev.}\ }\textbf {\bibinfo {volume} {D88}},\ \bibinfo {pages}
  {044054} (\bibinfo {year} {2013})},\ \Eprint {http://arxiv.org/abs/1306.4000}
  {arXiv:1306.4000 [hep-th]} \BibitemShut {NoStop}%
\bibitem [{\citenamefont {Dong}(2016)}]{Dong:2016fnf}%
  \BibitemOpen
  \bibfield  {author} {\bibinfo {author} {\bibfnamefont {X.}~\bibnamefont
  {Dong}},\ }\href {\doibase 10.1038/ncomms12472} {\bibfield  {journal}
  {\bibinfo  {journal} {Nature Commun.}\ }\textbf {\bibinfo {volume} {7}},\
  \bibinfo {pages} {12472} (\bibinfo {year} {2016})},\ \Eprint
  {http://arxiv.org/abs/1601.06788} {arXiv:1601.06788 [hep-th]} \BibitemShut
  {NoStop}%
\bibitem [{\citenamefont {Dong}(2014)}]{Dong:2013qoa}%
  \BibitemOpen
  \bibfield  {author} {\bibinfo {author} {\bibfnamefont {X.}~\bibnamefont
  {Dong}},\ }\href {\doibase 10.1007/JHEP01(2014)044} {\bibfield  {journal}
  {\bibinfo  {journal} {JHEP}\ }\textbf {\bibinfo {volume} {01}},\ \bibinfo
  {pages} {044} (\bibinfo {year} {2014})},\ \Eprint
  {http://arxiv.org/abs/1310.5713} {arXiv:1310.5713 [hep-th]} \BibitemShut
  {NoStop}%
\bibitem [{\citenamefont {Camps}(2014)}]{Camps:2013zua}%
  \BibitemOpen
  \bibfield  {author} {\bibinfo {author} {\bibfnamefont {J.}~\bibnamefont
  {Camps}},\ }\href {\doibase 10.1007/JHEP03(2014)070} {\bibfield  {journal}
  {\bibinfo  {journal} {JHEP}\ }\textbf {\bibinfo {volume} {03}},\ \bibinfo
  {pages} {070} (\bibinfo {year} {2014})},\ \Eprint
  {http://arxiv.org/abs/1310.6659} {arXiv:1310.6659 [hep-th]} \BibitemShut
  {NoStop}%
\bibitem [{\citenamefont {Miao}\ and\ \citenamefont
  {Guo}(2015)}]{Miao:2014nxa}%
  \BibitemOpen
  \bibfield  {author} {\bibinfo {author} {\bibfnamefont {R.-X.}\ \bibnamefont
  {Miao}}\ and\ \bibinfo {author} {\bibfnamefont {W.-z.}\ \bibnamefont {Guo}},\
  }\href {\doibase 10.1007/JHEP08(2015)031} {\bibfield  {journal} {\bibinfo
  {journal} {JHEP}\ }\textbf {\bibinfo {volume} {08}},\ \bibinfo {pages} {031}
  (\bibinfo {year} {2015})},\ \Eprint {http://arxiv.org/abs/1411.5579}
  {arXiv:1411.5579 [hep-th]} \BibitemShut {NoStop}%
\bibitem [{\citenamefont {Taylor}\ and\ \citenamefont
  {Woodhead}(2016)}]{Taylor:2016aoi}%
  \BibitemOpen
  \bibfield  {author} {\bibinfo {author} {\bibfnamefont {M.}~\bibnamefont
  {Taylor}}\ and\ \bibinfo {author} {\bibfnamefont {W.}~\bibnamefont
  {Woodhead}},\ }\href {\doibase 10.1007/JHEP08(2016)165} {\bibfield  {journal}
  {\bibinfo  {journal} {JHEP}\ }\textbf {\bibinfo {volume} {08}},\ \bibinfo
  {pages} {165} (\bibinfo {year} {2016})},\ \Eprint
  {http://arxiv.org/abs/1604.06808} {arXiv:1604.06808 [hep-th]} \BibitemShut
  {NoStop}%
\bibitem [{\citenamefont {Anastasiou}\ \emph
  {et~al.}(2018{\natexlab{a}})\citenamefont {Anastasiou}, \citenamefont
  {Araya},\ and\ \citenamefont {Olea}}]{Anastasiou:2018rla}%
  \BibitemOpen
  \bibfield  {author} {\bibinfo {author} {\bibfnamefont {G.}~\bibnamefont
  {Anastasiou}}, \bibinfo {author} {\bibfnamefont {I.~J.}\ \bibnamefont
  {Araya}}, \ and\ \bibinfo {author} {\bibfnamefont {R.}~\bibnamefont {Olea}},\
  }\href {\doibase 10.1103/PhysRevD.97.106015} {\bibfield  {journal} {\bibinfo
  {journal} {Phys. Rev.}\ }\textbf {\bibinfo {volume} {D97}},\ \bibinfo {pages}
  {106015} (\bibinfo {year} {2018}{\natexlab{a}})},\ \Eprint
  {http://arxiv.org/abs/1803.04990} {arXiv:1803.04990 [hep-th]} \BibitemShut
  {NoStop}%
\bibitem [{\citenamefont {Taylor}\ and\ \citenamefont
  {Too}(2020)}]{Taylor:2020uwf}%
  \BibitemOpen
  \bibfield  {author} {\bibinfo {author} {\bibfnamefont {M.}~\bibnamefont
  {Taylor}}\ and\ \bibinfo {author} {\bibfnamefont {L.}~\bibnamefont {Too}},\
  }\href {\doibase 10.1007/JHEP12(2020)050} {\bibfield  {journal} {\bibinfo
  {journal} {JHEP}\ }\textbf {\bibinfo {volume} {12}},\ \bibinfo {pages} {050}
  (\bibinfo {year} {2020})},\ \Eprint {http://arxiv.org/abs/2004.09568}
  {arXiv:2004.09568 [hep-th]} \BibitemShut {NoStop}%
\bibitem [{\citenamefont {Casini}\ and\ \citenamefont
  {Huerta}(2012)}]{Casini:2012ei}%
  \BibitemOpen
  \bibfield  {author} {\bibinfo {author} {\bibfnamefont {H.}~\bibnamefont
  {Casini}}\ and\ \bibinfo {author} {\bibfnamefont {M.}~\bibnamefont
  {Huerta}},\ }\href {\doibase 10.1103/PhysRevD.85.125016} {\bibfield
  {journal} {\bibinfo  {journal} {Phys. Rev.}\ }\textbf {\bibinfo {volume}
  {D85}},\ \bibinfo {pages} {125016} (\bibinfo {year} {2012})},\ \Eprint
  {http://arxiv.org/abs/1202.5650} {arXiv:1202.5650 [hep-th]} \BibitemShut
  {NoStop}%
\bibitem [{\citenamefont {Anastasiou}\ \emph
  {et~al.}(2020{\natexlab{a}})\citenamefont {Anastasiou}, \citenamefont
  {Moreno}, \citenamefont {Olea},\ and\ \citenamefont
  {Rivera-Betancour}}]{Anastasiou:2020smm}%
  \BibitemOpen
  \bibfield  {author} {\bibinfo {author} {\bibfnamefont {G.}~\bibnamefont
  {Anastasiou}}, \bibinfo {author} {\bibfnamefont {J.}~\bibnamefont {Moreno}},
  \bibinfo {author} {\bibfnamefont {R.}~\bibnamefont {Olea}}, \ and\ \bibinfo
  {author} {\bibfnamefont {D.}~\bibnamefont {Rivera-Betancour}},\ }\href
  {\doibase 10.1007/JHEP09(2020)173} {\bibfield  {journal} {\bibinfo  {journal}
  {JHEP}\ }\textbf {\bibinfo {volume} {09}},\ \bibinfo {pages} {173} (\bibinfo
  {year} {2020}{\natexlab{a}})},\ \Eprint {http://arxiv.org/abs/2002.06111}
  {arXiv:2002.06111 [hep-th]} \BibitemShut {NoStop}%
\bibitem [{\citenamefont {Anastasiou}\ \emph
  {et~al.}(2019{\natexlab{b}})\citenamefont {Anastasiou}, \citenamefont
  {Araya}, \citenamefont {Guijosa},\ and\ \citenamefont
  {Olea}}]{Anastasiou:2019ldc}%
  \BibitemOpen
  \bibfield  {author} {\bibinfo {author} {\bibfnamefont {G.}~\bibnamefont
  {Anastasiou}}, \bibinfo {author} {\bibfnamefont {I.~J.}\ \bibnamefont
  {Araya}}, \bibinfo {author} {\bibfnamefont {A.}~\bibnamefont {Guijosa}}, \
  and\ \bibinfo {author} {\bibfnamefont {R.}~\bibnamefont {Olea}},\ }\href
  {\doibase 10.1007/JHEP10(2019)221} {\bibfield  {journal} {\bibinfo  {journal}
  {JHEP}\ }\textbf {\bibinfo {volume} {10}},\ \bibinfo {pages} {221} (\bibinfo
  {year} {2019}{\natexlab{b}})},\ \Eprint {http://arxiv.org/abs/1908.11447}
  {arXiv:1908.11447 [hep-th]} \BibitemShut {NoStop}%
\bibitem [{\citenamefont {Henningson}\ and\ \citenamefont
  {Skenderis}(1998)}]{Henningson:1998gx}%
  \BibitemOpen
  \bibfield  {author} {\bibinfo {author} {\bibfnamefont {M.}~\bibnamefont
  {Henningson}}\ and\ \bibinfo {author} {\bibfnamefont {K.}~\bibnamefont
  {Skenderis}},\ }\href {\doibase 10.1088/1126-6708/1998/07/023} {\bibfield
  {journal} {\bibinfo  {journal} {JHEP}\ }\textbf {\bibinfo {volume} {07}},\
  \bibinfo {pages} {023} (\bibinfo {year} {1998})},\ \Eprint
  {http://arxiv.org/abs/hep-th/9806087} {arXiv:hep-th/9806087 [hep-th]}
  \BibitemShut {NoStop}%
\bibitem [{\citenamefont {Gibbons}\ and\ \citenamefont
  {Hawking}(1977)}]{Gibbons:1976ue}%
  \BibitemOpen
  \bibfield  {author} {\bibinfo {author} {\bibfnamefont {G.~W.}\ \bibnamefont
  {Gibbons}}\ and\ \bibinfo {author} {\bibfnamefont {S.~W.}\ \bibnamefont
  {Hawking}},\ }\href {\doibase 10.1103/PhysRevD.15.2752} {\bibfield  {journal}
  {\bibinfo  {journal} {Phys. Rev. D}\ }\textbf {\bibinfo {volume} {15}},\
  \bibinfo {pages} {2752} (\bibinfo {year} {1977})}\BibitemShut {NoStop}%
\bibitem [{\citenamefont {Emparan}\ \emph {et~al.}(1999)\citenamefont
  {Emparan}, \citenamefont {Johnson},\ and\ \citenamefont
  {Myers}}]{Emparan:1999pm}%
  \BibitemOpen
  \bibfield  {author} {\bibinfo {author} {\bibfnamefont {R.}~\bibnamefont
  {Emparan}}, \bibinfo {author} {\bibfnamefont {C.~V.}\ \bibnamefont
  {Johnson}}, \ and\ \bibinfo {author} {\bibfnamefont {R.~C.}\ \bibnamefont
  {Myers}},\ }\href {\doibase 10.1103/PhysRevD.60.104001} {\bibfield  {journal}
  {\bibinfo  {journal} {Phys. Rev.}\ }\textbf {\bibinfo {volume} {D60}},\
  \bibinfo {pages} {104001} (\bibinfo {year} {1999})},\ \Eprint
  {http://arxiv.org/abs/hep-th/9903238} {arXiv:hep-th/9903238 [hep-th]}
  \BibitemShut {NoStop}%
\bibitem [{\citenamefont {Balasubramanian}\ and\ \citenamefont
  {Kraus}(1999)}]{Balasubramanian:1999re}%
  \BibitemOpen
  \bibfield  {author} {\bibinfo {author} {\bibfnamefont {V.}~\bibnamefont
  {Balasubramanian}}\ and\ \bibinfo {author} {\bibfnamefont {P.}~\bibnamefont
  {Kraus}},\ }\href {\doibase 10.1007/s002200050764} {\bibfield  {journal}
  {\bibinfo  {journal} {Commun. Math. Phys.}\ }\textbf {\bibinfo {volume}
  {208}},\ \bibinfo {pages} {413} (\bibinfo {year} {1999})},\ \Eprint
  {http://arxiv.org/abs/hep-th/9902121} {arXiv:hep-th/9902121 [hep-th]}
  \BibitemShut {NoStop}%
\bibitem [{\citenamefont {Kraus}\ \emph {et~al.}(1999)\citenamefont {Kraus},
  \citenamefont {Larsen},\ and\ \citenamefont {Siebelink}}]{Kraus:1999di}%
  \BibitemOpen
  \bibfield  {author} {\bibinfo {author} {\bibfnamefont {P.}~\bibnamefont
  {Kraus}}, \bibinfo {author} {\bibfnamefont {F.}~\bibnamefont {Larsen}}, \
  and\ \bibinfo {author} {\bibfnamefont {R.}~\bibnamefont {Siebelink}},\ }\href
  {\doibase 10.1016/S0550-3213(99)00549-0} {\bibfield  {journal} {\bibinfo
  {journal} {Nucl. Phys.}\ }\textbf {\bibinfo {volume} {B563}},\ \bibinfo
  {pages} {259} (\bibinfo {year} {1999})},\ \Eprint
  {http://arxiv.org/abs/hep-th/9906127} {arXiv:hep-th/9906127 [hep-th]}
  \BibitemShut {NoStop}%
\bibitem [{\citenamefont {Papadimitriou}\ and\ \citenamefont
  {Skenderis}(2005)}]{Papadimitriou:2005ii}%
  \BibitemOpen
  \bibfield  {author} {\bibinfo {author} {\bibfnamefont {I.}~\bibnamefont
  {Papadimitriou}}\ and\ \bibinfo {author} {\bibfnamefont {K.}~\bibnamefont
  {Skenderis}},\ }\href {\doibase 10.1088/1126-6708/2005/08/004} {\bibfield
  {journal} {\bibinfo  {journal} {JHEP}\ }\textbf {\bibinfo {volume} {08}},\
  \bibinfo {pages} {004} (\bibinfo {year} {2005})},\ \Eprint
  {http://arxiv.org/abs/hep-th/0505190} {arXiv:hep-th/0505190 [hep-th]}
  \BibitemShut {NoStop}%
\bibitem [{\citenamefont {Witten}(2018)}]{Witten:2018lgb}%
  \BibitemOpen
  \bibfield  {author} {\bibinfo {author} {\bibfnamefont {E.}~\bibnamefont
  {Witten}},\ }\href@noop {} {\  (\bibinfo {year} {2018})},\ \Eprint
  {http://arxiv.org/abs/1805.11559} {arXiv:1805.11559 [hep-th]} \BibitemShut
  {NoStop}%
\bibitem [{\citenamefont {Aros}\ \emph
  {et~al.}(2000{\natexlab{a}})\citenamefont {Aros}, \citenamefont {Contreras},
  \citenamefont {Olea}, \citenamefont {Troncoso},\ and\ \citenamefont
  {Zanelli}}]{Aros:1999id}%
  \BibitemOpen
  \bibfield  {author} {\bibinfo {author} {\bibfnamefont {R.}~\bibnamefont
  {Aros}}, \bibinfo {author} {\bibfnamefont {M.}~\bibnamefont {Contreras}},
  \bibinfo {author} {\bibfnamefont {R.}~\bibnamefont {Olea}}, \bibinfo {author}
  {\bibfnamefont {R.}~\bibnamefont {Troncoso}}, \ and\ \bibinfo {author}
  {\bibfnamefont {J.}~\bibnamefont {Zanelli}},\ }\href {\doibase
  10.1103/PhysRevLett.84.1647} {\bibfield  {journal} {\bibinfo  {journal}
  {Phys. Rev. Lett.}\ }\textbf {\bibinfo {volume} {84}},\ \bibinfo {pages}
  {1647} (\bibinfo {year} {2000}{\natexlab{a}})},\ \Eprint
  {http://arxiv.org/abs/gr-qc/9909015} {arXiv:gr-qc/9909015} \BibitemShut
  {NoStop}%
\bibitem [{\citenamefont {Aros}\ \emph
  {et~al.}(2000{\natexlab{b}})\citenamefont {Aros}, \citenamefont {Contreras},
  \citenamefont {Olea}, \citenamefont {Troncoso},\ and\ \citenamefont
  {Zanelli}}]{Aros:1999kt}%
  \BibitemOpen
  \bibfield  {author} {\bibinfo {author} {\bibfnamefont {R.}~\bibnamefont
  {Aros}}, \bibinfo {author} {\bibfnamefont {M.}~\bibnamefont {Contreras}},
  \bibinfo {author} {\bibfnamefont {R.}~\bibnamefont {Olea}}, \bibinfo {author}
  {\bibfnamefont {R.}~\bibnamefont {Troncoso}}, \ and\ \bibinfo {author}
  {\bibfnamefont {J.}~\bibnamefont {Zanelli}},\ }\href {\doibase
  10.1103/PhysRevD.62.044002} {\bibfield  {journal} {\bibinfo  {journal} {Phys.
  Rev. D}\ }\textbf {\bibinfo {volume} {62}},\ \bibinfo {pages} {044002}
  (\bibinfo {year} {2000}{\natexlab{b}})},\ \Eprint
  {http://arxiv.org/abs/hep-th/9912045} {arXiv:hep-th/9912045} \BibitemShut
  {NoStop}%
\bibitem [{\citenamefont {Olea}(2005)}]{Olea:2005gb}%
  \BibitemOpen
  \bibfield  {author} {\bibinfo {author} {\bibfnamefont {R.}~\bibnamefont
  {Olea}},\ }\href {\doibase 10.1088/1126-6708/2005/06/023} {\bibfield
  {journal} {\bibinfo  {journal} {JHEP}\ }\textbf {\bibinfo {volume} {06}},\
  \bibinfo {pages} {023} (\bibinfo {year} {2005})},\ \Eprint
  {http://arxiv.org/abs/hep-th/0504233} {arXiv:hep-th/0504233 [hep-th]}
  \BibitemShut {NoStop}%
\bibitem [{\citenamefont {Olea}(2007)}]{Olea:2006vd}%
  \BibitemOpen
  \bibfield  {author} {\bibinfo {author} {\bibfnamefont {R.}~\bibnamefont
  {Olea}},\ }\href {\doibase 10.1088/1126-6708/2007/04/073} {\bibfield
  {journal} {\bibinfo  {journal} {JHEP}\ }\textbf {\bibinfo {volume} {04}},\
  \bibinfo {pages} {073} (\bibinfo {year} {2007})},\ \Eprint
  {http://arxiv.org/abs/hep-th/0610230} {arXiv:hep-th/0610230 [hep-th]}
  \BibitemShut {NoStop}%
\bibitem [{\citenamefont {Miskovic}\ and\ \citenamefont
  {Olea}(2009)}]{Miskovic:2009bm}%
  \BibitemOpen
  \bibfield  {author} {\bibinfo {author} {\bibfnamefont {O.}~\bibnamefont
  {Miskovic}}\ and\ \bibinfo {author} {\bibfnamefont {R.}~\bibnamefont
  {Olea}},\ }\href {\doibase 10.1103/PhysRevD.79.124020} {\bibfield  {journal}
  {\bibinfo  {journal} {Phys. Rev.}\ }\textbf {\bibinfo {volume} {D79}},\
  \bibinfo {pages} {124020} (\bibinfo {year} {2009})},\ \Eprint
  {http://arxiv.org/abs/0902.2082} {arXiv:0902.2082 [hep-th]} \BibitemShut
  {NoStop}%
\bibitem [{\citenamefont {Anastasiou}\ \emph
  {et~al.}(2020{\natexlab{b}})\citenamefont {Anastasiou}, \citenamefont
  {Miskovic}, \citenamefont {Olea},\ and\ \citenamefont
  {Papadimitriou}}]{Anastasiou:2020zwc}%
  \BibitemOpen
  \bibfield  {author} {\bibinfo {author} {\bibfnamefont {G.}~\bibnamefont
  {Anastasiou}}, \bibinfo {author} {\bibfnamefont {O.}~\bibnamefont
  {Miskovic}}, \bibinfo {author} {\bibfnamefont {R.}~\bibnamefont {Olea}}, \
  and\ \bibinfo {author} {\bibfnamefont {I.}~\bibnamefont {Papadimitriou}},\
  }\href {\doibase 10.1007/JHEP08(2020)061} {\bibfield  {journal} {\bibinfo
  {journal} {JHEP}\ }\textbf {\bibinfo {volume} {08}},\ \bibinfo {pages} {061}
  (\bibinfo {year} {2020}{\natexlab{b}})},\ \Eprint
  {http://arxiv.org/abs/2003.06425} {arXiv:2003.06425 [hep-th]} \BibitemShut
  {NoStop}%
\bibitem [{\citenamefont {{Anderson}}(2000)}]{2000math.....11051A}%
  \BibitemOpen
  \bibfield  {author} {\bibinfo {author} {\bibfnamefont {M.~T.}\ \bibnamefont
  {{Anderson}}},\ }\href@noop {} {\bibfield  {journal} {\bibinfo  {journal}
  {arXiv Mathematics e-prints}\ ,\ \bibinfo {eid} {math/0011051}} (\bibinfo
  {year} {2000})},\ \Eprint {http://arxiv.org/abs/math/0011051}
  {arXiv:math/0011051 [math.DG]} \BibitemShut {NoStop}%
\bibitem [{\citenamefont {{Albin}}(2005)}]{2005math......4161A}%
  \BibitemOpen
  \bibfield  {author} {\bibinfo {author} {\bibfnamefont {P.}~\bibnamefont
  {{Albin}}},\ }\href@noop {} {\bibfield  {journal} {\bibinfo  {journal} {arXiv
  Mathematics e-prints}\ ,\ \bibinfo {eid} {math/0504161}} (\bibinfo {year}
  {2005})},\ \Eprint {http://arxiv.org/abs/math/0504161} {arXiv:math/0504161
  [math.DG]} \BibitemShut {NoStop}%
\bibitem [{\citenamefont {Jatkar}\ \emph {et~al.}(2014)\citenamefont {Jatkar},
  \citenamefont {Kofinas}, \citenamefont {Miskovic},\ and\ \citenamefont
  {Olea}}]{Jatkar:2014npa}%
  \BibitemOpen
  \bibfield  {author} {\bibinfo {author} {\bibfnamefont {D.~P.}\ \bibnamefont
  {Jatkar}}, \bibinfo {author} {\bibfnamefont {G.}~\bibnamefont {Kofinas}},
  \bibinfo {author} {\bibfnamefont {O.}~\bibnamefont {Miskovic}}, \ and\
  \bibinfo {author} {\bibfnamefont {R.}~\bibnamefont {Olea}},\ }\href {\doibase
  10.1103/PhysRevD.89.124010} {\bibfield  {journal} {\bibinfo  {journal} {Phys.
  Rev. D}\ }\textbf {\bibinfo {volume} {89}},\ \bibinfo {pages} {124010}
  (\bibinfo {year} {2014})},\ \Eprint {http://arxiv.org/abs/1404.1411}
  {arXiv:1404.1411 [hep-th]} \BibitemShut {NoStop}%
\bibitem [{\citenamefont {Arenas-Henriquez}\ \emph {et~al.}(2017)\citenamefont
  {Arenas-Henriquez}, \citenamefont {Miskovic},\ and\ \citenamefont
  {Olea}}]{Arenas-Henriquez:2017xnr}%
  \BibitemOpen
  \bibfield  {author} {\bibinfo {author} {\bibfnamefont {G.}~\bibnamefont
  {Arenas-Henriquez}}, \bibinfo {author} {\bibfnamefont {O.}~\bibnamefont
  {Miskovic}}, \ and\ \bibinfo {author} {\bibfnamefont {R.}~\bibnamefont
  {Olea}},\ }\href {\doibase 10.1007/JHEP11(2017)128} {\bibfield  {journal}
  {\bibinfo  {journal} {JHEP}\ }\textbf {\bibinfo {volume} {11}},\ \bibinfo
  {pages} {128} (\bibinfo {year} {2017})},\ \Eprint
  {http://arxiv.org/abs/1710.08512} {arXiv:1710.08512 [hep-th]} \BibitemShut
  {NoStop}%
\bibitem [{\citenamefont {Arenas-Henriquez}\ \emph {et~al.}(2019)\citenamefont
  {Arenas-Henriquez}, \citenamefont {Mann}, \citenamefont {Miskovic},\ and\
  \citenamefont {Olea}}]{Arenas-Henriquez:2019rph}%
  \BibitemOpen
  \bibfield  {author} {\bibinfo {author} {\bibfnamefont {G.}~\bibnamefont
  {Arenas-Henriquez}}, \bibinfo {author} {\bibfnamefont {R.~B.}\ \bibnamefont
  {Mann}}, \bibinfo {author} {\bibfnamefont {O.}~\bibnamefont {Miskovic}}, \
  and\ \bibinfo {author} {\bibfnamefont {R.}~\bibnamefont {Olea}},\ }\href
  {\doibase 10.1103/PhysRevD.100.064038} {\bibfield  {journal} {\bibinfo
  {journal} {Phys. Rev. D}\ }\textbf {\bibinfo {volume} {100}},\ \bibinfo
  {pages} {064038} (\bibinfo {year} {2019})},\ \Eprint
  {http://arxiv.org/abs/1905.10840} {arXiv:1905.10840 [hep-th]} \BibitemShut
  {NoStop}%
\bibitem [{\citenamefont {Giribet}\ \emph {et~al.}(2018)\citenamefont
  {Giribet}, \citenamefont {Miskovic}, \citenamefont {Olea},\ and\
  \citenamefont {Rivera-Betancour}}]{Giribet:2018hck}%
  \BibitemOpen
  \bibfield  {author} {\bibinfo {author} {\bibfnamefont {G.}~\bibnamefont
  {Giribet}}, \bibinfo {author} {\bibfnamefont {O.}~\bibnamefont {Miskovic}},
  \bibinfo {author} {\bibfnamefont {R.}~\bibnamefont {Olea}}, \ and\ \bibinfo
  {author} {\bibfnamefont {D.}~\bibnamefont {Rivera-Betancour}},\ }\href
  {\doibase 10.1103/PhysRevD.98.044046} {\bibfield  {journal} {\bibinfo
  {journal} {Phys. Rev. D}\ }\textbf {\bibinfo {volume} {98}},\ \bibinfo
  {pages} {044046} (\bibinfo {year} {2018})},\ \Eprint
  {http://arxiv.org/abs/1806.11075} {arXiv:1806.11075 [hep-th]} \BibitemShut
  {NoStop}%
\bibitem [{\citenamefont {Giribet}\ \emph {et~al.}(2020)\citenamefont
  {Giribet}, \citenamefont {Miskovic}, \citenamefont {Olea},\ and\
  \citenamefont {Rivera-Betancour}}]{Giribet:2020aks}%
  \BibitemOpen
  \bibfield  {author} {\bibinfo {author} {\bibfnamefont {G.}~\bibnamefont
  {Giribet}}, \bibinfo {author} {\bibfnamefont {O.}~\bibnamefont {Miskovic}},
  \bibinfo {author} {\bibfnamefont {R.}~\bibnamefont {Olea}}, \ and\ \bibinfo
  {author} {\bibfnamefont {D.}~\bibnamefont {Rivera-Betancour}},\ }\href
  {\doibase 10.1103/PhysRevD.101.064046} {\bibfield  {journal} {\bibinfo
  {journal} {Phys. Rev. D}\ }\textbf {\bibinfo {volume} {101}},\ \bibinfo
  {pages} {064046} (\bibinfo {year} {2020})},\ \Eprint
  {http://arxiv.org/abs/2001.09459} {arXiv:2001.09459 [hep-th]} \BibitemShut
  {NoStop}%
\bibitem [{\citenamefont {Giribet}\ \emph {et~al.}()\citenamefont {Giribet},
  \citenamefont {Miskovic}, \citenamefont {Olea},\ and\ \citenamefont
  {Parra-Cisterna}}]{GMOPodd}%
  \BibitemOpen
  \bibfield  {author} {\bibinfo {author} {\bibfnamefont {G.}~\bibnamefont
  {Giribet}}, \bibinfo {author} {\bibfnamefont {O.}~\bibnamefont {Miskovic}},
  \bibinfo {author} {\bibfnamefont {R.}~\bibnamefont {Olea}}, \ and\ \bibinfo
  {author} {\bibfnamefont {Y.}~\bibnamefont {Parra-Cisterna}},\ }\href@noop {}
  {}\bibinfo {note} {In preparation}\BibitemShut {NoStop}%
\bibitem [{\citenamefont {Anastasiou}\ \emph
  {et~al.}(2018{\natexlab{b}})\citenamefont {Anastasiou}, \citenamefont
  {Araya},\ and\ \citenamefont {Olea}}]{Anastasiou:2017xjr}%
  \BibitemOpen
  \bibfield  {author} {\bibinfo {author} {\bibfnamefont {G.}~\bibnamefont
  {Anastasiou}}, \bibinfo {author} {\bibfnamefont {I.~J.}\ \bibnamefont
  {Araya}}, \ and\ \bibinfo {author} {\bibfnamefont {R.}~\bibnamefont {Olea}},\
  }\href {\doibase 10.1103/PhysRevD.97.106011} {\bibfield  {journal} {\bibinfo
  {journal} {Phys. Rev.}\ }\textbf {\bibinfo {volume} {D97}},\ \bibinfo {pages}
  {106011} (\bibinfo {year} {2018}{\natexlab{b}})},\ \Eprint
  {http://arxiv.org/abs/1712.09099} {arXiv:1712.09099 [hep-th]} \BibitemShut
  {NoStop}%
\bibitem [{\citenamefont {Kastikainen}(2020)}]{Kastikainen:2020auf}%
  \BibitemOpen
  \bibfield  {author} {\bibinfo {author} {\bibfnamefont {J.}~\bibnamefont
  {Kastikainen}},\ }\href {\doibase 10.1088/1361-6382/abac44} {\bibfield
  {journal} {\bibinfo  {journal} {Class. Quant. Grav.}\ }\textbf {\bibinfo
  {volume} {37}},\ \bibinfo {pages} {195010} (\bibinfo {year} {2020})},\
  \Eprint {http://arxiv.org/abs/2006.02803} {arXiv:2006.02803 [hep-th]}
  \BibitemShut {NoStop}%
\bibitem [{\citenamefont {Anastasiou}\ \emph
  {et~al.}(2018{\natexlab{c}})\citenamefont {Anastasiou}, \citenamefont
  {Araya}, \citenamefont {Arias},\ and\ \citenamefont
  {Olea}}]{Anastasiou:2018mfk}%
  \BibitemOpen
  \bibfield  {author} {\bibinfo {author} {\bibfnamefont {G.}~\bibnamefont
  {Anastasiou}}, \bibinfo {author} {\bibfnamefont {I.~J.}\ \bibnamefont
  {Araya}}, \bibinfo {author} {\bibfnamefont {C.}~\bibnamefont {Arias}}, \ and\
  \bibinfo {author} {\bibfnamefont {R.}~\bibnamefont {Olea}},\ }\href {\doibase
  10.1007/JHEP08(2018)136} {\bibfield  {journal} {\bibinfo  {journal} {JHEP}\
  }\textbf {\bibinfo {volume} {08}},\ \bibinfo {pages} {136} (\bibinfo {year}
  {2018}{\natexlab{c}})},\ \Eprint {http://arxiv.org/abs/1806.10708}
  {arXiv:1806.10708 [hep-th]} \BibitemShut {NoStop}%
\bibitem [{\citenamefont {Mohammadi~Mozaffar}\ \emph
  {et~al.}(2016)\citenamefont {Mohammadi~Mozaffar}, \citenamefont {Mollabashi},
  \citenamefont {Sheikh-Jabbari},\ and\ \citenamefont
  {Vahidinia}}]{Mozaffar:2016hmg}%
  \BibitemOpen
  \bibfield  {author} {\bibinfo {author} {\bibfnamefont {M.~R.}\ \bibnamefont
  {Mohammadi~Mozaffar}}, \bibinfo {author} {\bibfnamefont {A.}~\bibnamefont
  {Mollabashi}}, \bibinfo {author} {\bibfnamefont {M.~M.}\ \bibnamefont
  {Sheikh-Jabbari}}, \ and\ \bibinfo {author} {\bibfnamefont {M.~H.}\
  \bibnamefont {Vahidinia}},\ }\href {\doibase 10.1103/PhysRevD.94.046002}
  {\bibfield  {journal} {\bibinfo  {journal} {Phys. Rev. D}\ }\textbf {\bibinfo
  {volume} {94}},\ \bibinfo {pages} {046002} (\bibinfo {year} {2016})},\
  \Eprint {http://arxiv.org/abs/1603.05713} {arXiv:1603.05713 [hep-th]}
  \BibitemShut {NoStop}%
\bibitem [{\citenamefont {Camps}\ and\ \citenamefont
  {Kelly}(2015)}]{Camps:2014voa}%
  \BibitemOpen
  \bibfield  {author} {\bibinfo {author} {\bibfnamefont {J.}~\bibnamefont
  {Camps}}\ and\ \bibinfo {author} {\bibfnamefont {W.~R.}\ \bibnamefont
  {Kelly}},\ }\href {\doibase 10.1007/JHEP03(2015)061} {\bibfield  {journal}
  {\bibinfo  {journal} {JHEP}\ }\textbf {\bibinfo {volume} {03}},\ \bibinfo
  {pages} {061} (\bibinfo {year} {2015})},\ \Eprint
  {http://arxiv.org/abs/1412.4093} {arXiv:1412.4093 [hep-th]} \BibitemShut
  {NoStop}%
\bibitem [{\citenamefont {Miao}(2015)}]{Miao:2015iba}%
  \BibitemOpen
  \bibfield  {author} {\bibinfo {author} {\bibfnamefont {R.-X.}\ \bibnamefont
  {Miao}},\ }\href {\doibase 10.1007/JHEP10(2015)049} {\bibfield  {journal}
  {\bibinfo  {journal} {JHEP}\ }\textbf {\bibinfo {volume} {10}},\ \bibinfo
  {pages} {049} (\bibinfo {year} {2015})},\ \Eprint
  {http://arxiv.org/abs/1503.05538} {arXiv:1503.05538 [hep-th]} \BibitemShut
  {NoStop}%
\bibitem [{\citenamefont {Camps}(2016)}]{Camps:2016gfs}%
  \BibitemOpen
  \bibfield  {author} {\bibinfo {author} {\bibfnamefont {J.}~\bibnamefont
  {Camps}},\ }\href {\doibase 10.1007/JHEP09(2016)139} {\bibfield  {journal}
  {\bibinfo  {journal} {JHEP}\ }\textbf {\bibinfo {volume} {09}},\ \bibinfo
  {pages} {139} (\bibinfo {year} {2016})},\ \Eprint
  {http://arxiv.org/abs/1605.08588} {arXiv:1605.08588 [hep-th]} \BibitemShut
  {NoStop}%
\bibitem [{\citenamefont {Bhattacharyya}\ and\ \citenamefont
  {Sharma}(2014)}]{Bhattacharyya:2014yga}%
  \BibitemOpen
  \bibfield  {author} {\bibinfo {author} {\bibfnamefont {A.}~\bibnamefont
  {Bhattacharyya}}\ and\ \bibinfo {author} {\bibfnamefont {M.}~\bibnamefont
  {Sharma}},\ }\href {\doibase 10.1007/JHEP10(2014)130} {\bibfield  {journal}
  {\bibinfo  {journal} {JHEP}\ }\textbf {\bibinfo {volume} {10}},\ \bibinfo
  {pages} {130} (\bibinfo {year} {2014})},\ \Eprint
  {http://arxiv.org/abs/1405.3511} {arXiv:1405.3511 [hep-th]} \BibitemShut
  {NoStop}%
\bibitem [{\citenamefont {Bueno}\ \emph {et~al.}(2018)\citenamefont {Bueno},
  \citenamefont {Cano},\ and\ \citenamefont {Ruip\'erez}}]{Bueno:2018xqc}%
  \BibitemOpen
  \bibfield  {author} {\bibinfo {author} {\bibfnamefont {P.}~\bibnamefont
  {Bueno}}, \bibinfo {author} {\bibfnamefont {P.~A.}\ \bibnamefont {Cano}}, \
  and\ \bibinfo {author} {\bibfnamefont {A.}~\bibnamefont {Ruip\'erez}},\
  }\href {\doibase 10.1007/JHEP03(2018)150} {\bibfield  {journal} {\bibinfo
  {journal} {JHEP}\ }\textbf {\bibinfo {volume} {03}},\ \bibinfo {pages} {150}
  (\bibinfo {year} {2018})},\ \Eprint {http://arxiv.org/abs/1802.00018}
  {arXiv:1802.00018 [hep-th]} \BibitemShut {NoStop}%
\bibitem [{\citenamefont {Bueno}\ \emph
  {et~al.}(2020{\natexlab{c}})\citenamefont {Bueno}, \citenamefont {Camps},\
  and\ \citenamefont {L\'opez}}]{Bueno:2020uxs}%
  \BibitemOpen
  \bibfield  {author} {\bibinfo {author} {\bibfnamefont {P.}~\bibnamefont
  {Bueno}}, \bibinfo {author} {\bibfnamefont {J.}~\bibnamefont {Camps}}, \ and\
  \bibinfo {author} {\bibfnamefont {A.~V.}\ \bibnamefont {L\'opez}},\
  }\href@noop {} {\  (\bibinfo {year} {2020}{\natexlab{c}})},\ \Eprint
  {http://arxiv.org/abs/2012.14033} {arXiv:2012.14033 [hep-th]} \BibitemShut
  {NoStop}%
\bibitem [{\citenamefont {Schwimmer}\ and\ \citenamefont
  {Theisen}(2008)}]{Schwimmer:2008yh}%
  \BibitemOpen
  \bibfield  {author} {\bibinfo {author} {\bibfnamefont {A.}~\bibnamefont
  {Schwimmer}}\ and\ \bibinfo {author} {\bibfnamefont {S.}~\bibnamefont
  {Theisen}},\ }\href {\doibase 10.1016/j.nuclphysb.2008.04.015} {\bibfield
  {journal} {\bibinfo  {journal} {Nucl. Phys. B}\ }\textbf {\bibinfo {volume}
  {801}},\ \bibinfo {pages} {1} (\bibinfo {year} {2008})},\ \Eprint
  {http://arxiv.org/abs/0802.1017} {arXiv:0802.1017 [hep-th]} \BibitemShut
  {NoStop}%
\bibitem [{\citenamefont {Araya}\ \emph {et~al.}(2021)\citenamefont {Araya},
  \citenamefont {Edelstein}, \citenamefont {Sanchez}, \citenamefont
  {Rodriguez},\ and\ \citenamefont {Lopez}}]{Araya:2021atx}%
  \BibitemOpen
  \bibfield  {author} {\bibinfo {author} {\bibfnamefont {I.~J.}\ \bibnamefont
  {Araya}}, \bibinfo {author} {\bibfnamefont {J.~D.}\ \bibnamefont
  {Edelstein}}, \bibinfo {author} {\bibfnamefont {A.~R.}\ \bibnamefont
  {Sanchez}}, \bibinfo {author} {\bibfnamefont {D.~V.}\ \bibnamefont
  {Rodriguez}}, \ and\ \bibinfo {author} {\bibfnamefont {A.~V.}\ \bibnamefont
  {Lopez}},\ }\href@noop {} {\  (\bibinfo {year} {2021})},\ \Eprint
  {http://arxiv.org/abs/2108.01126} {arXiv:2108.01126 [hep-th]} \BibitemShut
  {NoStop}%
\bibitem [{\citenamefont {Imbimbo}\ \emph {et~al.}(2000)\citenamefont
  {Imbimbo}, \citenamefont {Schwimmer}, \citenamefont {Theisen},\ and\
  \citenamefont {Yankielowicz}}]{Imbimbo:1999bj}%
  \BibitemOpen
  \bibfield  {author} {\bibinfo {author} {\bibfnamefont {C.}~\bibnamefont
  {Imbimbo}}, \bibinfo {author} {\bibfnamefont {A.}~\bibnamefont {Schwimmer}},
  \bibinfo {author} {\bibfnamefont {S.}~\bibnamefont {Theisen}}, \ and\
  \bibinfo {author} {\bibfnamefont {S.}~\bibnamefont {Yankielowicz}},\ }\href
  {\doibase 10.1088/0264-9381/17/5/322} {\bibfield  {journal} {\bibinfo
  {journal} {Class. Quant. Grav.}\ }\textbf {\bibinfo {volume} {17}},\ \bibinfo
  {pages} {1129} (\bibinfo {year} {2000})},\ \Eprint
  {http://arxiv.org/abs/hep-th/9910267} {arXiv:hep-th/9910267} \BibitemShut
  {NoStop}%
\bibitem [{\citenamefont {de~Haro}\ \emph {et~al.}(2001)\citenamefont
  {de~Haro}, \citenamefont {Solodukhin},\ and\ \citenamefont
  {Skenderis}}]{deHaro:2000vlm}%
  \BibitemOpen
  \bibfield  {author} {\bibinfo {author} {\bibfnamefont {S.}~\bibnamefont
  {de~Haro}}, \bibinfo {author} {\bibfnamefont {S.~N.}\ \bibnamefont
  {Solodukhin}}, \ and\ \bibinfo {author} {\bibfnamefont {K.}~\bibnamefont
  {Skenderis}},\ }\href {\doibase 10.1007/s002200100381} {\bibfield  {journal}
  {\bibinfo  {journal} {Commun. Math. Phys.}\ }\textbf {\bibinfo {volume}
  {217}},\ \bibinfo {pages} {595} (\bibinfo {year} {2001})},\ \Eprint
  {http://arxiv.org/abs/hep-th/0002230} {arXiv:hep-th/0002230 [hep-th]}
  \BibitemShut {NoStop}%
\end{thebibliography}%
\label{biblio}

\end{document}